\documentclass[fleqn,usenatbib,pdflatex,iicol]{mnras}
\usepackage{url}
\usepackage{graphicx}	
\usepackage{pdflscape}
\usepackage{longtable}

\RequirePackage{color}

\newcommand{\msun}{M$_{\odot}$}
\newcommand{\lsun}{L$_{\odot}$}

\providecommand{\parallax}{\ensuremath{\varpi}}

\providecommand{\sigparallax}{\ensuremath{\sigma_{\varpi}}}
\providecommand{\fpu}{\ensuremath{\parallax/\sigparallax}}

\title[Post-AGB stars in Gaia DR3]{A census of Post-AGB stars in Gaia DR3: evidence for a substantial population of Galactic post-{\em RGB} stars}

\author[René D. Oudmaijer et al.]{
René D. Oudmaijer,$^{1}$\thanks{E-mail: r.d.oudmaijer@leeds.ac.uk}
Emma R. M. Jones,$^{1}$
and Miguel Vioque$^{2,3}$
\\
$^{1}$School of Physics \& Astronomy, University of Leeds, Woodhouse Lane, LS2 9JT, Leeds, UK\\
$^{2}$Joint ALMA Observatory, Alonso de Córdova 3107, Vitacura 763-0355, Santiago, Chile\\
$^{3}$National Radio Astronomy Observatory, 520 Edgemont Road, Charlottesville, VA 22903, USA}

\date{Accepted for publication in MNRAS Letters, 2022 July 31.}

\pubyear{2022}

\begin{document}
\label{firstpage}
\pagerange{\pageref{firstpage}--\pageref{lastpage}}
\maketitle

\begin{abstract}
This paper presents the first census of Galactic post-Asymptotic Giant Branch stars in the HR diagram.  We combined Gaia DR3 parallax-based distances  with extinction corrected integrated fluxes, and derived luminosities for a sample of 185 stars that had been proposed to be post-AGB stars in the literature. 
The luminosities allow us to create an HR diagram containing the largest number of post-AGB candidate objects to date.  A significant fraction of the objects fall outside the typical luminosity range as covered by theoretical evolutionary post-AGB tracks as well as observed for Planetary Nebula central stars.  These include massive evolved supergiants and lower luminosity objects. Here we highlight the fact that one third of the post-AGB candidates is underluminous and we identify these with the recently recognised class of post-Red Giant Branch objects  thought to be the result of binary evolution. 
\end{abstract}


\section{Introduction}
\label{sec:intro}

Prior to its final fate as a cooling White Dwarf, the Sun will be going through the Planetary Nebula (PN) phase after having moved from the Asymptotic Giant Branch (AGB) through the subsequent post-AGB (or pre-PN) stage. This scenario is believed to apply to most low- to intermediate mass stars \citep{Vanwinckel2003}. During the post-AGB phase, the star moves rapidly to the blue in the HR diagram. The phase is typically assumed to end when the central star has reached a surface temperature of 30,000 K, when the star can ionise its envelope and the object appears as a Planetary Nebula. 
Post-AGB evolutionary timescales depend on the stars' initial masses, and vary from $\sim$5000 to $\sim$200 years for stars with initial masses of 1 \msun \, and 4 M$_{\odot}$ \, respectively \citep{Schonberner1983,Blocker1995}. These are likely upper limits, as, for example, increased post-AGB mass loss rates shorten the transition \citep{Vanhoof1997, Bertolami2019}.  

These timescales are  extremely short
which has hampered the search and identification of post-AGB candidate stars, and their identification came relatively late. AFGL 2688 and AFGL 618 were among the first to be identified as such. Their then newly discovered infrared excess emission due to circumstellar dust betrayed a previous mass losing evolutionary phase \citep{Ney1975,Westbrook1975}.  Much progress has been made since and many more post-AGB candidates stars have been identified and proposed.  Of note are the compilations by \citet{Suarez2006}  and the Toru\'n catalogue by \citet{Szczerba2007}.  

A crucial parameter to parameterise the objects and to place them in an evolutionary context is their luminosity. As the evolution from the AGB to the PN stage occurs more-or-less horizontally in the Hertzsprung-Russell (HR) diagram,  the luminosity immediately sets the stars' birth mass, core mass and evolutionary age. Distances have been hard to come by however.
A general approach has been to assume a fixed luminosity for all objects, and use the observed integrated flux to then derive the distance. 
This was the fundamental approach in the first distance catalogue of post-AGB stars published by \citet{Vickers2015}. These authors grouped the  known post-AGB candidates into subpopulations which were assigned luminosities between 3,500 and 10,000 L$_{\odot}$. The distance was then determined using the ratio of this luminosity and the total integrated flux which they derived from the stars' Spectral Energy Distributions (SEDs). 

With the advent of the Gaia satellite \citep{Gaia2016,Gaia2021} the time is ripe to collate the best available distances for post-AGB stars, determine their luminosities and, where possible, position them in the HR-diagram. 
\citet{Hrivnak2020} presented Gaia DR2-derived distances for 12 objects, \cite{Parthasarathy2020} presented data for a further 8 objects, while 
in a recent paper, \citet{Kamath2022} 
study 31 Galactic post-AGB stars and their Gaia DR3 data. However, a global analysis is yet to appear.  Here we make use of the fact that \citet{Vickers2015} derived integrated fluxes for 350 post-AGB stars. In this paper, we  derive distances and luminosities for those objects that are present in the Gaia DR3 release and investigate their properties. This results in the most complete census of Galactic post-AGB stars to date.

\begin{figure}
\includegraphics[scale=0.17,angle=0]{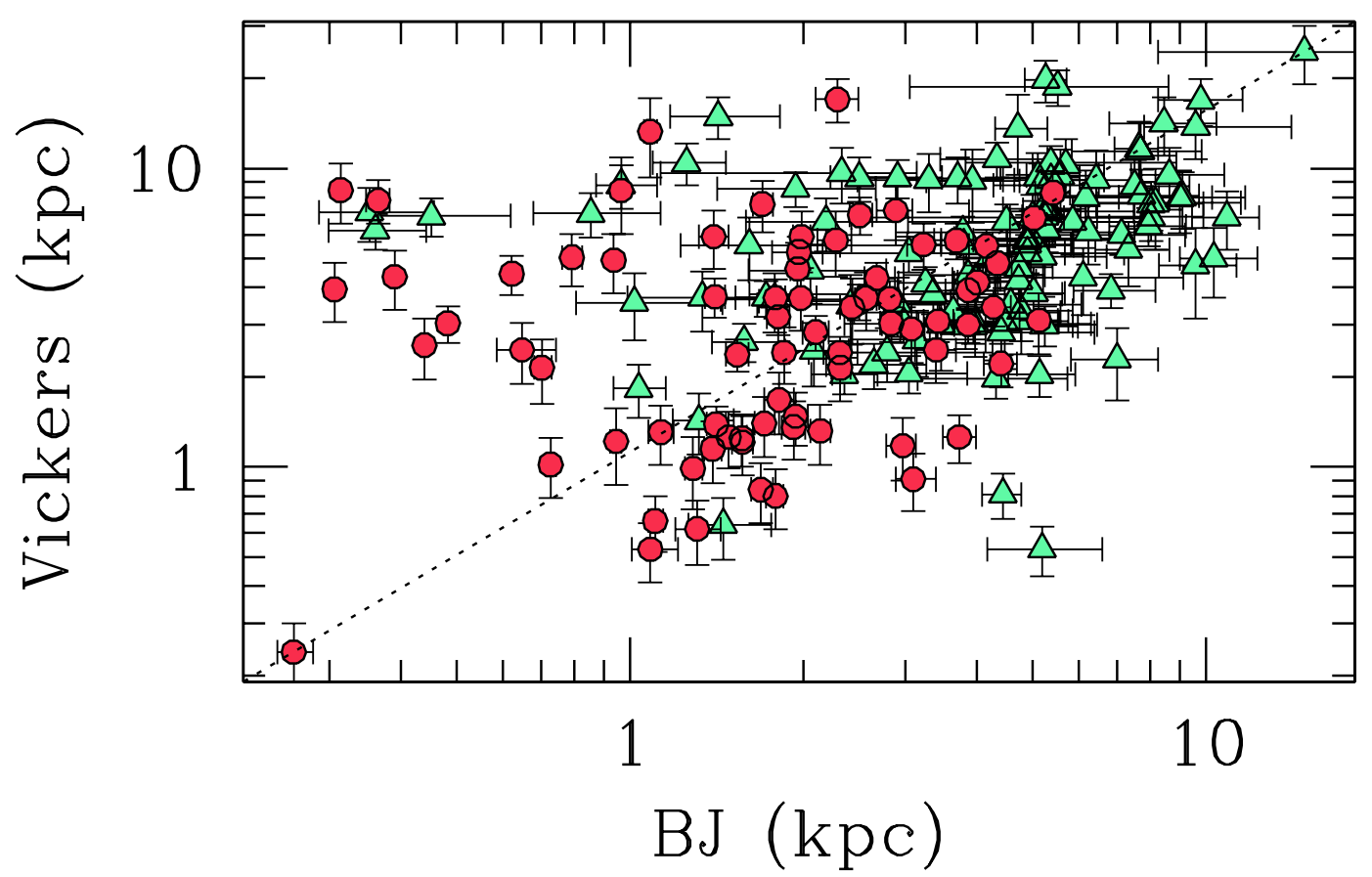}
\caption{The Bailer-Jones distances for the 185 objects with good parallaxes versus the Vickers distances.  
Stars with the best parallaxes ($\fpu > 10$) are denoted by the  circles, those with $1 < \fpu < 10$ by the  triangles. The correlation coefficient (in log-space) is 0.37. The dashed line represents the $x = y$ line. 
} 
\label{bjvickers}
\end{figure}
\section{Sample selection and data}
\label{sampledata}
We draw the sample from \citet{Vickers2015}. They  determined the integrated fluxes in a homogeneous manner for 209 ``likely" and 87 ``possible" post-AGB stars from the Toru\'n catalogue by \citet{Szczerba2007}, an additional 54  ``miscellaneous" objects  were suggested after that catalogue's publication. 
\citet{Vickers2015} determined the total integrated flux through fitting multiple black body functions to the SED for all 350 objects, after de-reddening the spectral energy distributions (SED) using interstellar extinction values presented by
\citet{Schlafly2011} and \citet{Arenou1992}.
Although this approach does not account for circumstellar reddening, it is to first order a good approximation as the re-emitted energy in the infrared is captured fully by the SED in case of spherical shells, while the approach was homogeneously applied to the entire sample. The uncertainties on the fluxes listed by \citet{Vickers2015} are typically 20\%. 

\begin{figure*}
\begin{center}
\includegraphics[scale=0.2,angle=0]{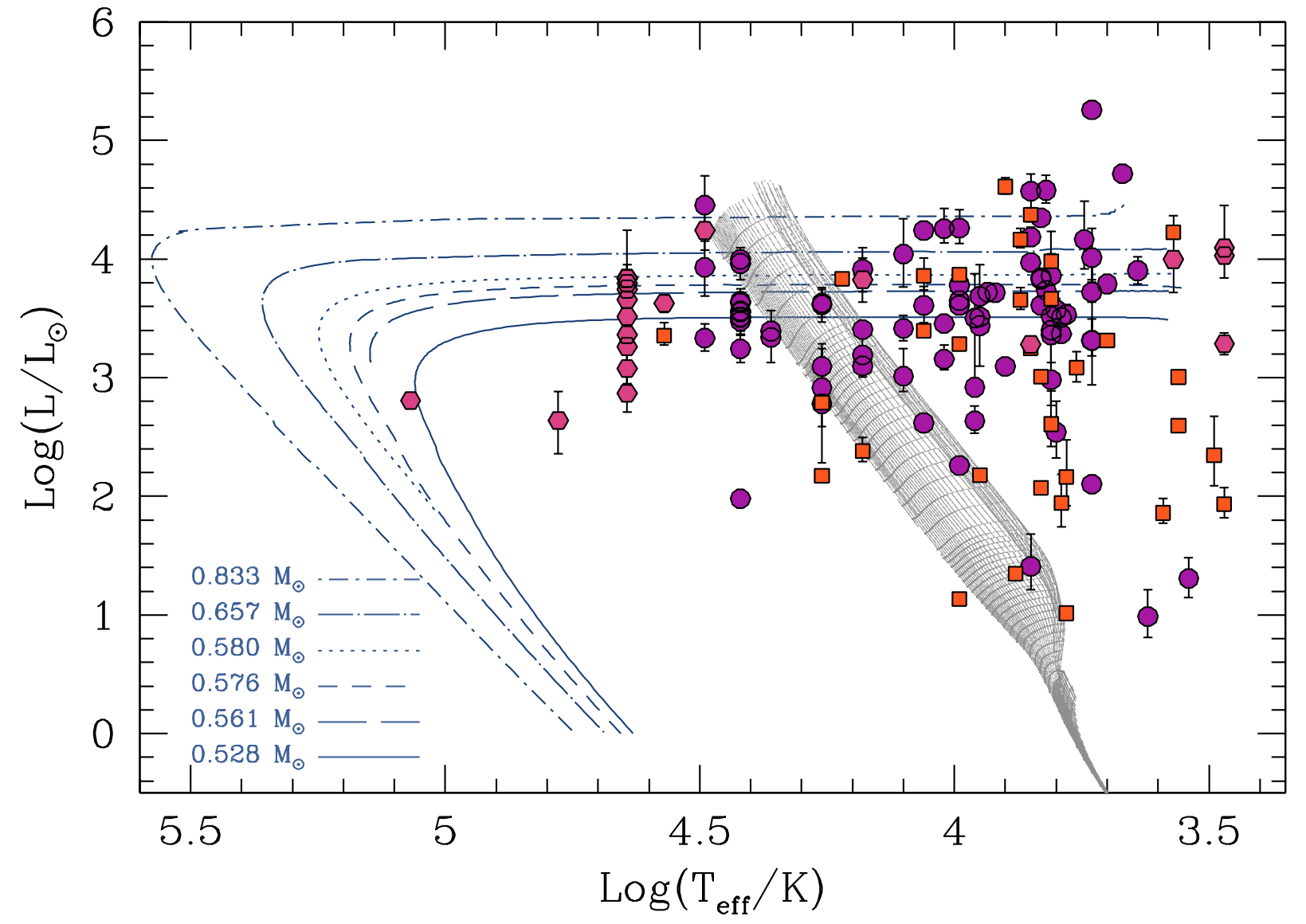}
\caption{HR diagram of the 134 post-AGB objects with both Gaia-based distances and temperature information available. The post-AGB evolutionary tracks are taken from \protect\citet{Bertolami2016} and computed for solar metallicity. The (core) masses are indicated in the bottom left, and correspond to intial masses from 1 - 4 \msun .  The Main Sequence is taken from the PARSEC tracks \protect\citep{Bressan2012}.
The ``likely" post-AGB  objects are denoted by the purple circles,  the ``possible" objects as  filled orange squares and the ``miscellaneous" post-AGB stars are represented by pink hexagons. The object classifications are those from \protect\citet{Vickers2015}.
} 
\label{hr}
\end{center}
\end{figure*}

150 of the 209 ``likely", 62/87 ``possible" and 37/54 miscellaneous post-AGB stars (so 249/350 in total) have Gaia DR3 parallaxes \citep{Gaia2021}. 
Distances based on parallax data can be determined in various ways, for high quality parallaxes it can be straightforwardly determined by inverting the parallax. For less well determined parallaxes, the larger errorbars result in notably a-symmetric errors once inverted and an accurate distance determination based on a simple inversion proves not possible. This can be alleviated by assuming an $a \, priori$ distribution of stars in the Milky Way;  \citet{BailerJones2021} used a Bayesian approach to obtain new distance estimates based on Gaia DR3 parallaxes.  
Two sets of distances were returned by Bailer-Jones et al. for DR3, the geometric distances which were solely based on the astrometry and the photometric distances which were additionally based on the stellar magnitudes and colours. As the latter method makes implicit assumptions about the nature of the objects, we proceed with the geometric distances. For objects which have $\fpu < 1$ or even a negative parallax, the Bailer-Jones values tend to converge to the prior, and distances thus derived are arguably not very useful.

249 out of the 350 Vickers post-AGB candidates have a DR3 parallax. Of those, 74 have very well determined parallaxes ($\fpu > 10$), 111 have good parallaxes ($1 < \fpu < 10$), while the remaining 64 objects have poorly determined parallaxes with $\fpu < 1$. 
We will continue in the following with the 185 objects with good parallaxes and use the Bailer-Jones distances throughout. We note that  the inverted parallaxes and Bailer-Jones distances are almost identical  for the highest quality parallaxes.

\subsection{Comparison with the Vickers distance catalogue}

It is an interesting exercise to see the impact Gaia has on our knowledge of post-AGB stars and their evolution. As noted above, distances of these objects were hard to come by and \citet{Vickers2015} produced the largest and arguably best distance catalogue for post-AGB stars at the time. Fig.~\ref{bjvickers} shows the Vickers distances against the Gaia-based distances. It appears there is hardly a correlation and indeed, a correlation coefficient of 0.37 is returned. Many objects are, often significantly, closer than the Vickers distances. This means that their Gaia-based luminosities are much lower than originally assumed. We are forced to conclude that the pre-Gaia distances were not a good indication of the true situation. This can in part be due to the underlying assumption that the objects are post-AGB stars. If they are not, then the originally assumed post-AGB luminosity would be wrong to begin with. We return to this notion later.

\section{Results}

\subsection{The HR Diagram}

In order to place the objects in an HR-diagram, we need both the luminosities of the objects and their temperature.  The stellar luminosities are computed by multiplying the, dereddened, integrated fluxes (in terms of luminosity per kiloparsec squared)  from \citet{Vickers2015} with the square of the Gaia-based distance (in kpc). 
Regarding the temperatures, we found spectral types for 167/249 objects,
the vast majority of those were sourced from the SIMBAD database.  For the spectral type to temperature conversion, we consulted the tables by \citet{Strazys1981}. As we will use the HR diagram only for illustrative purposes in this paper, we do not need precise temperatures. 

The resulting HR diagram is shown in Fig.~\ref{hr} which contains the 134 post-AGB stars for which we have both Gaia-based distances and temperatures based on the spectral types. For reference, the post-AGB evolutionary tracks \citep{Bertolami2016} are plotted. The core masses of the tracks range from 0.528 \msun, corresponding an initial mass of 1 \msun,  to a track for 0.833 \msun, corresponding to stars with an initial mass of 4 \msun. These tracks encompass the bulk of the objects that would be observable in their post-AGB phase at present; 
not many stars in the Galaxy with initial masses lower than 1 \msun \, will have evolved through the post-AGB phase, while their crossing times become so long that their circumstellar material will have long dispersed into the interstellar medium.  Objects with initial masses of 4 \msun \, have a post-AGB crossing time less than 100 years, a timescale that rapidly decreases with initial mass. There are not many high mass objects in general, but given the shorter  timescales, few higher mass objects will be observable as a post-AGB star. 

Most post-AGB candidate stars are indeed located along the theoretical evolutionary tracks. However,  many objects, including some with the smallest distance uncertainties, are located outside these most extreme tracks, both above (up to more than 10$^5$ \lsun) and below (as low at 10 \lsun) the lines.  We recall that the HR diagram only contains a sub-sample of objects, namely those with spectral types, allowing their positioning in the HR diagram. Let us now investigate the fuller sample of 185 objects with Gaia-based luminosities.

\subsection{The luminosities of post-AGB star candidates}

Fig.~\ref{lum1} shows a histogram of the luminosity of the objects from the ``likely", ``possible" and ``miscellaneous" categories and the total sample. All groups peak in the lower end of the luminosity range for post-AGB stars.
Using the mass-luminosity relationship by \cite{Herwig1998}, the peak luminosity translates into a mass that agrees with  the observed  peak mass of White Dwarf stars of 0.562 \msun\ reported by \citet{Bergeron1992} and which corresponds to $\sim$4665~\lsun . The values also agree with those observed for Planetary Nebula central stars before they enter the cooling track \citep[][based on Gaia data]{Gonzalez2021}.

However, the luminosities span a wide range from 10 to several 10$^{5}$ \lsun .  A significant fraction of the objects falls outside the \citet{Bertolami2016} post-AGB tracks for stars with initial masses of 1 - 4 \msun \ respectively. The two brightest objects are found to be IRAS 17163-3907, the central star of the Fried Egg nebula and HD 179821. Both are known massive evolved stars in their post-Red Supergiant phase \citep{Koumpia2020,Oudmaijer2009} and due to their similarity with post-AGB stars had often been included in such samples. Yet, the large number of underluminous objects is arguably the main eye-catching finding. These will be discussed below. 

\section{Discussion}

We combined Gaia DR3 data  with extinction corrected integrated fluxes, and derived luminosities for a large sample of objects that have been proposed to be post-AGB stars in the literature. Most objects have luminosities as expected for the class, yet many appear to be underluminous. Before discussing their nature, let us first investigate whether any observational effects have affected the results. 
 \newline
 

\noindent

\begin{figure}
\includegraphics[scale=0.16,angle=0]{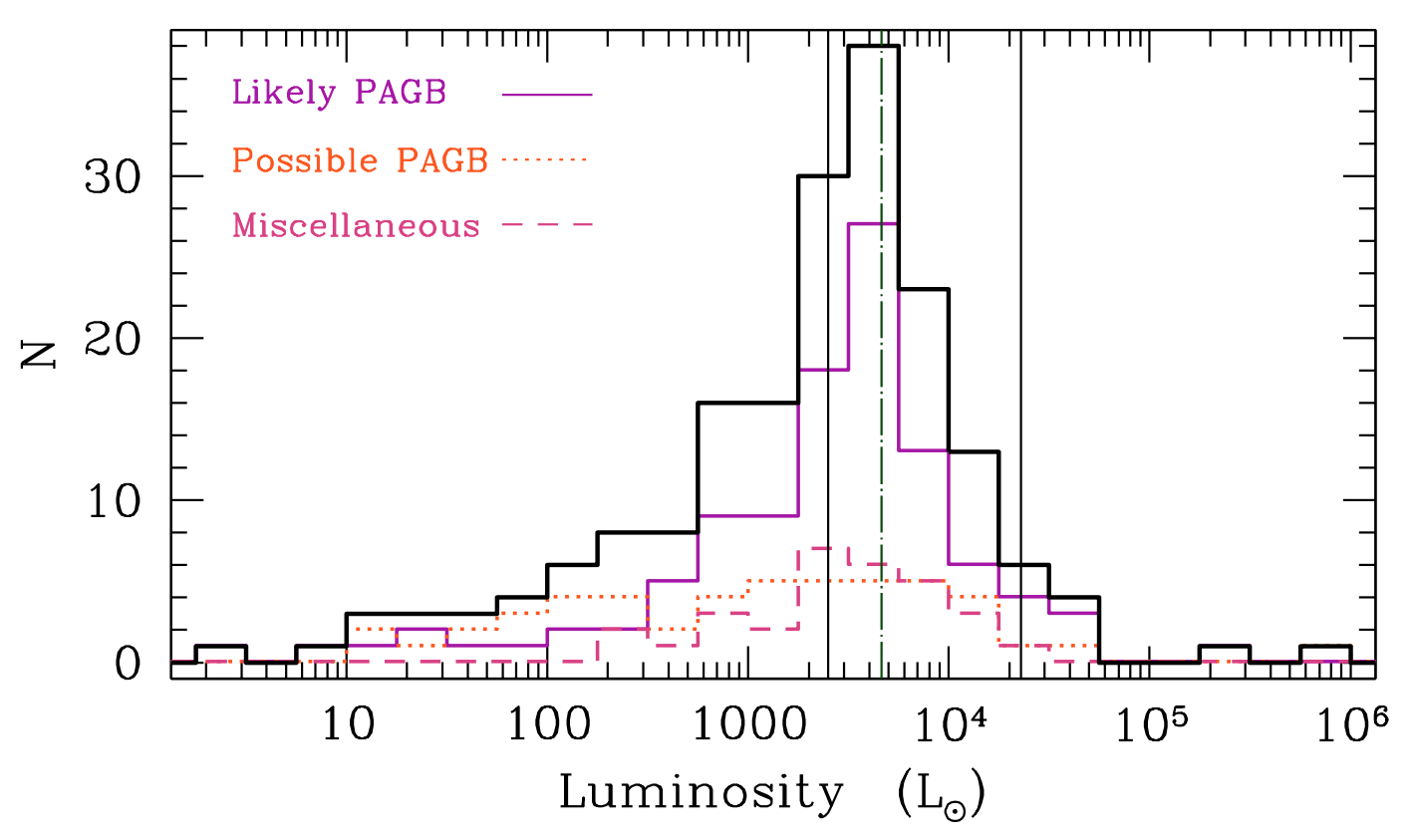}
\caption{
The Gaia-based luminosities of 185 post-AGB objects for  the various categories of post-AGB star as listed by \protect\citet{Vickers2015}, and the total.   
The solid vertical lines correspond to luminosities for 1 and 4 \msun \ objects respectively. The dashed vertical line indicates the typical luminosity observed for the central stars of Planetary Nebulae (see text).
} 
\label{lum1}
\end{figure}

{\it The luminosity determination:} 
The main errors on the luminosity determination are due to uncertainties in the distance and the flux determination for the objects themselves. To start with the former, the histogram bin sizes in Fig.~\ref{lum1} are larger than the errors on the stars' luminosities, while it can also be seen in the HR diagram that objects with the smallest errors on the luminosity are clearly below the canonical post-AGB tracks, strongly suggesting the observational uncertainties do not bias the results. 

Another uncertainty entering the computation of the luminosity is that of the determination of the total integrated flux. \citet{Vickers2015} report that the errors on their flux determination are of order 20\%.
As a check
on the accuracy of these authors'  global approach, we compare the total integrated fluxes as derived by them with the 31 objects published in \citet{Kamath2022}. The latter authors took into account the total extinction using known stellar parameters in contrast to \citet{Vickers2015} who only took into account interstellar extinction as outlined earlier. 
The literature values are on average 1.2 $\pm$ 0.4 larger than those computed by Vickers. 
Although there is a systematic difference, 
this can be due to the uncertainty and choice of the reddening  values \citep[cf.][]{Kamath2022}. 
Although the literature comparison suggests that Vickers' fluxes may have been a bit underestimated, we conclude that both the uncertainties on the fluxes as well as any corrections to these are too small to explain the extent to which a large number of objects is underluminous.

\smallskip

\begin{figure}
\includegraphics[scale=0.145,angle=0]{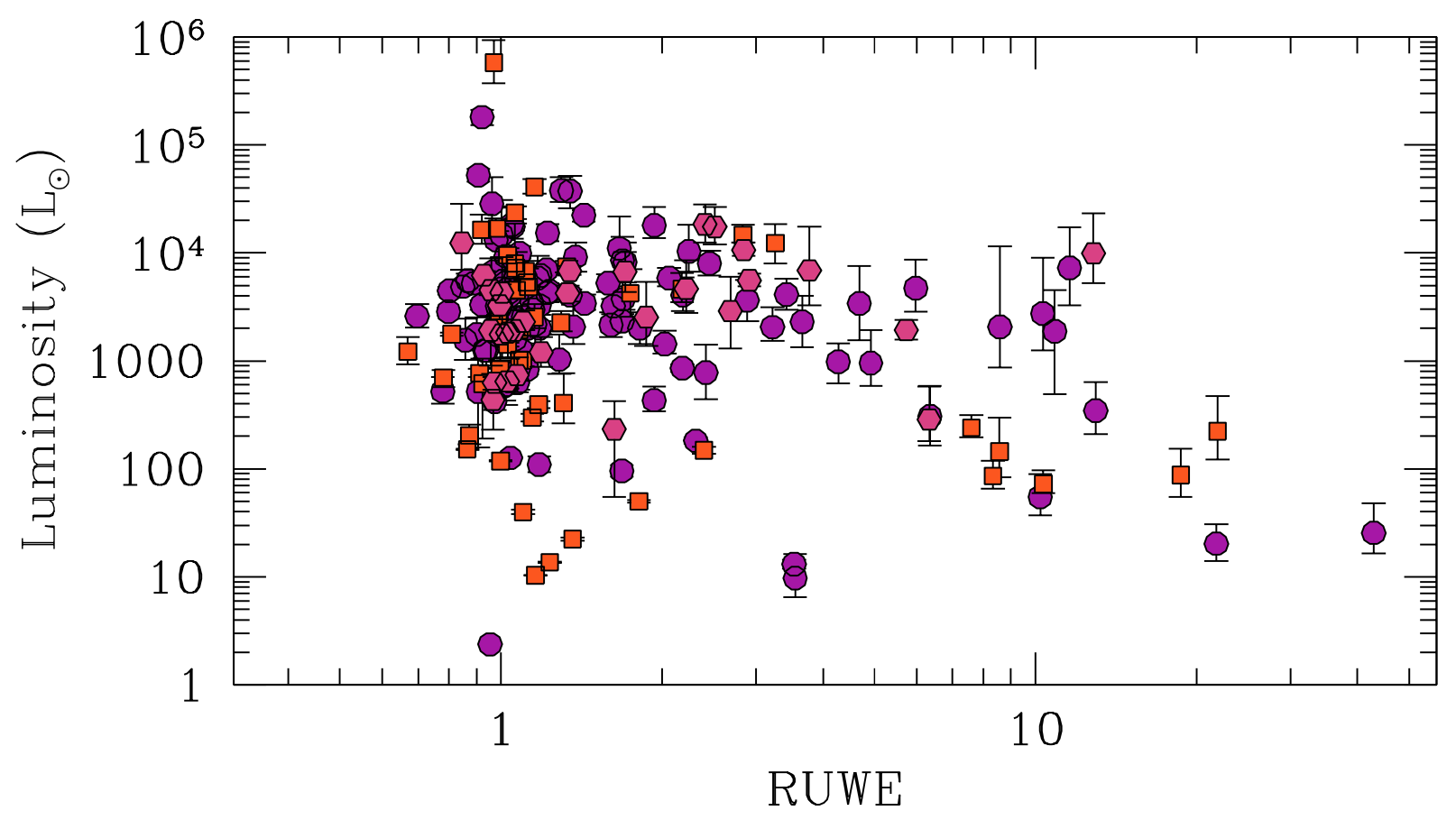}
\caption{The luminosity as function of RUWE parameter. 
The plot symbols are the same as in Fig.~\ref{hr}. 
} 
\label{qual}
\end{figure}

\noindent
{\it Effect of parallax quality on luminosities:}  Next to the uncertainties on the parallax  discussed above, the renormalised unit weight error (RUWE, 
\citealt{Lindegren2021}) is a powerful quality indicator of the Gaia astrometry. It measures the goodness-of-fit for the astrometric solution and is sensitive to the presence of extended emission or binary companions. 
It may be useful to point out that 
both AFGL 618 and AFGL 2688 which were mentioned earlier do not have a Gaia-based parallax measurement. These objects are very extended at optical wavelengths and determining their positions and thus parallaxes proved difficult. 
Fig.~\ref{qual} shows the luminosity as function of RUWE.  Many of the lowest luminosity stars have values far below the typical value of RUWE $<$ 1.4 which is normally taken as of good quality. This suggests that the RUWE parameter, or the circumstances that lead to its determination, is not responsible for the large number of under-luminous objects. 
The objects with the highest RUWE are among the faintest, closest, objects. We suspect that more distant, more luminous objects which would otherwise receive a high RUWE value,  would have their parallaxes more compromised and not be included in Gaia DR3. It appears that the parallax quality - as probed by the RUWE parameter - does not bias the results.
\newline

\noindent
{\it Vickers' classification of post-AGB nature:}
\citet{Vickers2015} 
classified objects as being a ``likely" or ``possible" post-AGB star 
and added a number of ``miscellaneous" post-AGB candidates. As Figures~\ref{hr} and \ref{lum1}  indicate, all three categories contain underluminous objects.
 The ``likely" group - the group with the best credentials to be a post-AGB object according to \citet{Vickers2015} 
 contains a large number of underluminous objects; 25 out of the 106 objects in the category are fainter than 1000 \lsun.  Almost half (22 out of 49) of the ``possible" group 
 appears to be underluminous, while 6 out of 30 ``miscellaneous" stars are underluminous. 
It would appear that the confidence in the classification of having a post-AGB nature did not have a significant effect on whether the objects are underluminous or not.

\subsection{The underluminous post-AGB stars: Galactic post-Red Giant Branch objects.}

This paper presents the first census of post-AGB candidate stars in the HR diagram. Here we highlight that many objects are placed well below the widely accepted post-AGB tracks. 53 objects out of the 185 stars for which luminosities could be derived are fainter than 1000 \lsun. This finding comes at the heels of \citet{Hrivnak2020} who studied 5 post-AGB stars and pointed out that 2 of these had (Gaia-DR2 based) luminosities of 1000-1500 \lsun. 
In addition, recently, \citet{Kamath2022} reported that 4 of their 31 post-AGB stars, selected to be both ``single" and being bright enough to have abundance analyses performed,  were  $<$ 1000 \lsun \ based on their Gaia DR3 derived luminosities.

Pre-Gaia, the most reliable  HR-diagram could only be constructed for extra-Galactic objects which were all at the same, known, distance. 
\citet{Kamath2016} found that about a third of the post-AGB candidate stars in the Magellanic Clouds were also underluminous, with values down to $\sim$100 \lsun . These authors pointed out that Red Giant Branch (RGB) stars that have similar luminosities as the objects under consideration do not have sufficiently large mass loss rates to give rise to observable infrared excess emission. Instead, they proposed the objects are the result of binary evolution. In particular, they suggest that the stars did evolve off the RGB after a recent Common Envelope  phase perhaps followed by a merger. The fraction of post-RGB candidates in their sample was in line with predictions from population synthesis models.  The
large number of lower luminosity stars  we find appear to be the Galactic counterparts of this new class of object, confirming  that these objects are numerous. Indeed, completing the evolutionary picture, post-RGB candidate stars have also been reported in Planetary Nebulae \citep{Hillwig2017,Jones2022}. 

It is intriguing that although our sample is less well-defined as the Magellanic Cloud sample, we arrive at a similar fraction of post-RGB objects. We do note however that the underluminous objects are predominately found at smaller distances. A properly defined, volume-limited, sample might well increase that fraction, but this would require a well-defined sample with well-understood observational biases, which is thus far lacking for the Galaxy.

Our analysis earlier on Gaia's RUWE parameter which can be used as an indication of binarity  \citep{Belokurov2020},  demonstrated that many under-luminous objects do not have large RUWE values. In that sense, they can be considered single or as an unresolved close binary (e.g. \citealt{Chornay2021}). 
Although this is very much circumstantial, the ``non- or close binary" nature can be seen as consistent with the notion that they are the result of Common Envelope Evolution.


\section{Concluding Remarks}

We have presented the first Gaia census of post-AGB stars. From a master sample of 350 objects, we were able to determine luminosities for 185,  and place 134 of these, the largest such sample ever, in the HR diagram. We find the following.

\begin{itemize}
    
\item

Our principal conclusion is that although many objects are very consistent with the post-AGB evolutionary tracks, we find that 53, almost a third of this total, are underluminous and unlikely to have a post-AGB nature. 

\item
Instead, we identify these as the Galactic counterparts of the class of post-Red Giant Branch stars recently discovered in the Magellanic Clouds. As such we report the discovery of the first sample of Galactic post-RGB stars.

\item Current models suggest that these post-RGB objects are the result of binary interactions and possibly mergers. As these objects turn out to be very numerous, they will pose significant challenges to, and provide crucial new data to inform, binary star evolution and population synthesis models. 
\end{itemize}

Finally, moving to the extremes, like \citet{Kamath2022} we too find objects with luninosities down to $\sim$10 \lsun. This would be too faint to be explained by RGB progenitors which are expected to be brighter than $\sim$100 \lsun . On the other extreme, some of the brightest, overluminous, objects ($\sim 10^{4-5}$~\lsun), are known massive evolved stars in their post-Red Supergiant phase \citep{Oudmaijer2009}. The sample is likely to harbour more such examples.  In future work we aim to  study these samples in more detail.


\section*{Acknowledgments} 

This work has made use of data from the European Space Agency (ESA) mission
{\it Gaia} (\url{https://www.cosmos.esa.int/gaia}), processed by the {\it Gaia}
Data Processing and Analysis Consortium (DPAC,
\url{https://www.cosmos.esa.int/web/gaia/dpac/consortium}). Funding for the DPAC
has been provided by national institutions, in particular the institutions
participating in the {\it Gaia} Multilateral Agreement.

\section*{Data availability}
The catalogue of post-AGB stars with parallax information, their astrometric data and derived parameters used in this article are available in its online supplementary material and in Table \ref{table_final}.

\bibliography{biblio-u1}

\clearpage
\onecolumn
\renewcommand{\arraystretch}{1.2}

\begin{landscape}
Table 1: The 249 sources from \citet{Vickers2015} for which parallax information is available. The table lists the astrometric data and the stellar parameters used in this article. Distances from \citet{BailerJones2021}. Last column indicates the post-AGB likelihood classification from \citet{Szczerba2007} and \citet{Vickers2015}: `1' for the \textit{likely} category, `2' for the \textit{possible} category, and `3' for the \textit{miscellaneous} category.
\begin{longtable}{llrrlrrrrrr}
\hline
Gaia DR3 source id & SIMBAD name & RA & DEC & SpType & T\textsubscript{eff} & Parallax & RUWE & Distance & Luminosity & Vickers\\
 &  & h:m:s & deg:m:s & & log([K]) & [mas] & & [kpc] & [L$_{\odot}$] & category\\
\hline\label{table_final}
565507868441719424 & IRAS 01005+7910 & 01:04:45.5 & +79:26:47 & B2 & 4.36 & $0.241\pm{0.018}$ & 1.037 & $3.68^{+0.26}_{-0.22}$ & $2470^{+360}_{-290}$ & 1 \\
2351623413515105920 & BPS CS22946-0005 & 01:16:52.9 & -22:12:09 & B & 4.26 & $-0.036\pm{0.034}$ & 1.017 & $13.8^{+3.4}_{-4.3}$ & $1240^{+680}_{-640}$ & 2 \\
4715635535640762240 & LB 3193 & 01:18:53.3 & -61:55:41 &  &  & $0.137\pm{0.024}$ & 0.7787 & $5.71^{+0.93}_{-0.71}$ & $520^{+180}_{-120}$ & 1 \\
532078488709487360 & IRAS 01259+6823 & 01:29:33.3 & +68:39:15 & F5Ie & 3.81 & $0.62\pm{0.14}$ & 1.309 & $1.78^{+0.68}_{-0.35}$ & $410^{+370}_{-140}$ & 2 \\
4686479751449676032 & LB 3219 & 01:30:22.9 & -73:03:33 & B & 4.26 & $-0.067\pm{0.024}$ & 0.9284 & $29^{+11}_{-9}$ & $49000^{+45000}_{-26000}$ & 1 \\
351149177434709760 & IRAS 01427+4633 & 01:45:47.0 & +46:49:02 & F2III & 3.85 & $0.293\pm{0.017}$ & 1.058 & $3.08^{+0.16}_{-0.13}$ & $1910^{+210}_{-160}$ & 3 \\
459182413984008448 & IRAS 02143+5852 & 02:17:57.7 & +59:05:53 & F7Ie & 3.79 & $1.36\pm{0.29}$ & 18.69 & $0.85^{+0.28}_{-0.17}$ & $87^{+66}_{-32}$ & 2 \\
513671461473684352 & IRAS Z02229+6208 & 02:26:41.6 & +62:21:23 & K0 & 3.64 & $0.381\pm{0.06}$ & 2.454 & $2.35^{+0.34}_{-0.29}$ & $8000^{+2400}_{-1800}$ & 1 \\
433515788197481984 & IRAS 02528+4350 & 02:56:11.3 & +44:02:54 & A0e & 3.99 & $2.541\pm{0.019}$ & 1.234 & $0.3899^{+0.0027}_{-0.0032}$ & $13.68^{+0.19}_{-0.22}$ & 2 \\
3303343395568710016 & IRAS 03507+1115 & 03:53:28.9 & +11:24:21 & M7 & 3.47 & $3.84\pm{0.27}$ & 2.844 & $0.260^{+0.021}_{-0.017}$ & $10700^{+1800}_{-1400}$ & 3 \\
471908436438311680 & IRAS 04215+6000 & 04:25:50.8 & +60:07:12 & W9 & 4.643 & $0.153\pm{0.016}$ & 1.1 & $5.23^{+0.68}_{-0.44}$ & $2300^{+630}_{-370}$ & 3 \\
173086700992466688 & IRAS 04296+3429 & 04:32:57.0 & +34:36:11 & G0Ia & 3.76 & $-0.38\pm{0.17}$ & 5.758 & $5.0^{+2.1}_{-1.1}$ & $6200^{+6200}_{-2500}$ & 1 \\
255225480926107392 & IRAS 05040+4820 & 05:07:50.4 & +48:24:09 & A4Ia & 3.935 & $0.322\pm{0.014}$ & 1.02 & $2.84^{+0.12}_{-0.13}$ & $5260^{+470}_{-460}$ & 1 \\
3238918336374596864 & IRAS 05089+0459 & 05:11:36.2 & +05:03:26 & M3I & 3.49 & $0.75\pm{0.35}$ & 21.91 & $1.34^{+0.61}_{-0.35}$ & $220^{+250}_{-100}$ & 2 \\
3388902129107252992 & IRAS 05113+1347 & 05:14:07.7 & +13:50:29 & G5I & 3.7 & $-0.01\pm{0.15}$ & 6.663 & $4.6^{+3.8}_{-1.3}$ & $2600^{+6000}_{-1300}$ & 1 \\
3422437684728294528 & IRAS 05140+2851 & 05:17:15.5 & +28:54:20 & F0 & 3.87 & $0.894\pm{0.079}$ & 2.834 & $1.08^{+0.12}_{-0.08}$ & $14600^{+3500}_{-2100}$ & 2 \\
2968265509022275840 & IRAS 05208-2035 & 05:22:59.5 & -20:32:52 &  &  & $0.687\pm{0.03}$ & 2.184 & $1.403^{+0.059}_{-0.053}$ & $854^{+73}_{-64}$ & 1 \\
4758015524139610880 & CPD-61 455 & 05:23:33.7 & -60:55:28 & B1 & 4.42 & $0.694\pm{0.035}$ & 1.683 & $1.395^{+0.085}_{-0.077}$ & $95^{+12}_{-10}$ & 1 \\
2902505745786910080 & IRAS F05338-3051 & 05:35:44.2 & -30:49:35 & G5 & 3.7 & $0.522\pm{0.016}$ & 1.094 & $1.850^{+0.043}_{-0.044}$ & $2050^{+100}_{-100}$ & 2 \\
3334854780347915520 & IRAS 05341+0852 & 05:36:55.2 & +08:54:08 & F6I & 3.8 & $0.51\pm{0.19}$ & 12.96 & $2.06^{+0.72}_{-0.45}$ & $350^{+290}_{-140}$ & 1 \\
3336558507975208448 & IRAS 05381+1012 & 05:40:57.1 & +10:14:25 & G2I & 3.73 & $1.038\pm{0.016}$ & 1.042 & $0.937^{+0.010}_{-0.011}$ & $126.4^{+2.8}_{-2.9}$ & 1 \\
994259335315643520 & IRAS 06338+5333 & 06:37:52.2 & +53:31:02 & F7IV: & 3.79 & $0.268\pm{0.024}$ & 1.622 & $3.40^{+0.28}_{-0.28}$ & $3210^{+550}_{-500}$ & 1 \\
3105987960396950784 & IRAS 06530-0213 & 06:55:31.7 & -02:17:27 & F5Ia & 3.81 & $0.241\pm{0.074}$ & 3.657 & $3.8^{+1.2}_{-0.9}$ & $2300^{+1700}_{-1000}$ & 1 \\
3159640386918214528 & IRAS 07008+1050 & 07:03:39.8 & +10:46:12 & A0 & 3.99 & $0.518\pm{0.03}$ & 1.348 & $1.81^{+0.11}_{-0.11}$ & $4070^{+510}_{-460}$ & 1 \\
3108327343185135872 & IRAS 07131-0147 & 07:15:41.8 & -01:52:40 & M5III & 3.47 & $0.79\pm{0.13}$ & 8.333 & $1.25^{+0.22}_{-0.16}$ & $86^{+32}_{-20}$ & 2 \\
5617989266685365120 & IRAS 07140-2321 & 07:16:08.3 & -23:27:03 & F5 & 3.81 & $0.178\pm{0.012}$ & 1.029 & $5.12^{+0.40}_{-0.38}$ & $9500^{+1600}_{-1400}$ & 2 \\
3156171118495247360 & IRAS 07134+1005 & 07:16:10.3 & +09:59:47 & F5I & 3.81 & $0.454\pm{0.024}$ & 0.9215 & $2.10^{+0.11}_{-0.11}$ & $3300^{+360}_{-330}$ & 1 \\
3032030620730261376 & IRAS 07227-1320 & 07:25:03.0 & -13:26:20 & M1I & 3.56 & $0.489\pm{0.021}$ & 1.176 & $1.982^{+0.074}_{-0.067}$ & $393^{+30}_{-26}$ & 2 \\
5620444471847839232 & IRAS 07253-2001 & 07:27:33.0 & -20:07:21 & F2I & 3.85 & $2.45\pm{0.52}$ & 42.89 & $0.45^{+0.17}_{-0.09}$ & $25.6^{+22.4}_{-9.2}$ & 1 \\
3151417586128916864 & IRAS 07430+1115 & 07:45:51.4 & +11:08:20 & M2I & 3.54 & $3.06\pm{0.5}$ & 21.8 & $0.361^{+0.082}_{-0.061}$ & $20.3^{+10.2}_{-6.3}$ & 1 \\
5597822402371118336 & IRAS 07577-2806 & 07:59:46.5 & -28:14:54 &  &  & $0.78\pm{0.16}$ & 10.21 & $1.42^{+0.40}_{-0.25}$ & $55^{+35}_{-17}$ & 1 \\
\hline
\end{longtable}
\end{landscape}

\begin{landscape}
Table 1 (continued).
\begin{longtable}{llrrlrrrrrr}
\hline
Gaia DR3 source id & SIMBAD name & RA & DEC & SpType & T\textsubscript{eff} & Parallax & RUWE & Distance & Luminosity & Vickers\\
 &  & h:m:s & deg:m:s & & log([K]) & [mas] & & [kpc] & [L$_{\odot}$] & category\\
\hline
5698817012142459136 & IRAS 08005-2356 & 08:02:40.8 & -24:04:44 & F5Iae & 3.81 & $-0.257\pm{0.083}$ & 5.902 & $9.2^{+2.1}_{-1.7}$ & $54000^{+28000}_{-18000}$ & 1 \\
5545800762036628736 & IRAS 08057-3417 & 08:07:40.0 & -34:26:04 &  &  & $0.258\pm{0.025}$ & 2.062 & $3.86^{+0.44}_{-0.33}$ & $5800^{+1400}_{-1000}$ & 1 \\
5520238967817034880 & IRAS 08143-4406 & 08:16:03.1 & -44:16:05 & F8I & 3.78 & $0.238\pm{0.018}$ & 1.433 & $4.16^{+0.41}_{-0.28}$ & $3400^{+710}_{-440}$ & 1 \\
5707613169577769600 & IRAS 08187-1905 & 08:20:57.2 & -19:15:04 & F6Ib/II & 3.8 & $0.288\pm{0.033}$ & 1.695 & $3.26^{+0.39}_{-0.34}$ & $3730^{+950}_{-740}$ & 1 \\
5540178478053582592 & IRAS 08242-3828 & 08:26:03.7 & -38:38:47 &  &  & $0.496\pm{0.052}$ & 1.155 & $2.09^{+0.31}_{-0.23}$ & $2540^{+810}_{-520}$ & 2 \\
5277809440015969792 & IRAS 08275-6206 & 08:28:24.5 & -62:16:20 &  &  & $0.496\pm{0.014}$ & 1.073 & $1.952^{+0.050}_{-0.057}$ & $629^{+32}_{-36}$ & 1 \\
5515266327706463616 & IRAS 08281-4850 & 08:29:40.6 & -49:00:03 & F0I & 3.87 & $-0.137\pm{0.069}$ & 6.044 & $11.4^{+3.7}_{-2.7}$ & $8400^{+6200}_{-3500}$ & 1 \\
5521628033275348480 & IRAS 08351-4634 & 08:36:45.4 & -46:44:49 &  &  & $0.505\pm{0.05}$ & 0.9962 & $1.96^{+0.19}_{-0.20}$ & $840^{+170}_{-160}$ & 2 \\
5409069172514684416 & IRAS 09394-4909 & 09:41:14.0 & -49:22:46 &  &  & $0.148\pm{0.041}$ & 2.407 & $5.2^{+1.2}_{-0.9}$ & $18000^{+10000}_{-6000}$ & 3 \\
5462428643590805248 & IRAS 10158-2844 & 10:18:07.5 & -28:59:31 & B9 & 4.02 & $0.711\pm{0.098}$ & 1.935 & $1.45^{+0.31}_{-0.19}$ & $18000^{+8600}_{-4300}$ & 1 \\
5254793942926363392 & IRAS 10214-6017 & 10:23:09.4 & -60:32:43 & O7 & 4.57 & $0.223\pm{0.023}$ & 1.332 & $4.20^{+0.36}_{-0.37}$ & $4240^{+750}_{-710}$ & 3 \\
5351904394753372672 & IRAS 10256-5628 & 10:27:35.3 & -56:44:19 & F5I & 3.81 & $-0.09\pm{0.33}$ & 6.799 & $3.8^{+2.1}_{-1.2}$ & $2800^{+3800}_{-1500}$ & 1 \\
5351069693654349952 & IRAS 10456-5712 & 10:47:38.5 & -57:28:03 & M0 & 3.57 & $0.9\pm{0.026}$ & 0.984 & $1.104^{+0.035}_{-0.029}$ & $16800^{+1100}_{-900}$ & 2 \\
5241806275407841664 & IRAS 11000-6153 & 11:02:04.4 & -62:09:44 & F2III & 3.85 & $0.221\pm{0.015}$ & 1.062 & $4.40^{+0.29}_{-0.29}$ & $23700^{+3200}_{-3100}$ & 2 \\
5337582534294739456 & IRAS 11159-5954 & 11:18:07.3 & -60:10:39 &  &  & $-0.29\pm{0.13}$ & 2.095 & $7.0^{+2.3}_{-1.5}$ & $4500^{+3500}_{-1700}$ & 2 \\
5237007177683569536 & IRAS 11201-6545 & 11:22:18.8 & -66:01:49 & A3Ie & 3.95 & $0.24\pm{0.13}$ & 10.32 & $4.5^{+3.6}_{-1.5}$ & $2700^{+6200}_{-1500}$ & 1 \\
5335866849446446080 & IRAS 11339-6004 & 11:36:20.9 & -60:20:53 &  &  & $0.08\pm{0.3}$ & 1.026 & $4.7^{+2.8}_{-2.0}$ & $800^{+1200}_{-500}$ & 1 \\
5335709477519159936 & IRAS 11353-6037 & 11:37:43.1 & -60:53:50 & B5Ie & 4.18 & $0.169\pm{0.013}$ & 1.101 & $5.42^{+0.38}_{-0.38}$ & $2550^{+370}_{-340}$ & 1 \\
5332682659443937664 & IRAS 11381-6401 & 11:40:31.7 & -64:18:26 &  &  & $-0.09\pm{0.2}$ & 0.9462 & $6.0^{+2.6}_{-1.7}$ & $4700^{+4900}_{-2200}$ & 1 \\
5343168568718268800 & IRAS 11385-5517 & 11:40:58.9 & -55:34:27 & B8 & 4.06 & $0.545\pm{0.02}$ & 1.054 & $1.788^{+0.057}_{-0.066}$ & $17400^{+1200}_{-1200}$ & 1 \\
5335675087769798272 & IRAS 11387-6113 & 11:41:08.7 & -61:30:19 & A3Ie & 3.95 & $0.185\pm{0.018}$ & 1.179 & $5.02^{+0.64}_{-0.47}$ & $3250^{+880}_{-580}$ & 1 \\
3589047952995134720 & IRAS 11472-0800 & 11:49:48.0 & -08:17:21 &  &  & $0.11\pm{0.084}$ & 4.678 & $6.1^{+3.0}_{-2.0}$ & $3400^{+4200}_{-1800}$ & 1 \\
3492184311482349824 & EC 11507-2253 & 11:53:15.9 & -23:09:53 & B5 & 4.18 & $-0.001\pm{0.038}$ & 0.9783 & $11.0^{+4.8}_{-2.4}$ & $330^{+350}_{-130}$ & 1 \\
5335102207846402176 & IRAS 11531-6111 & 11:55:37.8 & -61:28:16 & B8Ie & 4.06 & $0.158\pm{0.018}$ & 0.9739 & $5.27^{+0.46}_{-0.43}$ & $416^{+75}_{-65}$ & 1 \\
5332853912685160064 & IRAS 11544-6408 & 11:56:57.2 & -64:25:16 &  &  & $-0.55\pm{0.84}$ & 1.23 & $4.6^{+2.5}_{-1.5}$ & $4200^{+5900}_{-2300}$ & 1 \\
3920735495441657728 & BD+13 2491 & 12:07:10.9 & +12:59:07 & B9 & 4.02 & $0.597\pm{0.027}$ & 0.798 & $1.567^{+0.065}_{-0.064}$ & $2860^{+240}_{-230}$ & 1 \\
6076326701687231872 & IRAS 12175-5338 & 12:20:15.0 & -53:55:29 & A7I & 3.9 & $-0.009\pm{0.066}$ & 3.651 & $8.7^{+3.8}_{-2.2}$ & $20000^{+21000}_{-9000}$ & 1 \\
3469106382752903168 & CD-31 9638 & 12:20:45.0 & -32:33:25 & A2II/III & 3.96 & $0.367\pm{0.021}$ & 1.12 & $2.57^{+0.14}_{-0.14}$ & $831^{+91}_{-88}$ & 1 \\
6130448958959242240 & IRAS 12222-4652 & 12:24:53.4 & -47:09:09 & F4 & 3.82 & $0.17\pm{0.03}$ & 1.298 & $5.13^{+0.79}_{-0.60}$ & $38000^{+13000}_{-8000}$ & 1 \\
6053621271182676352 & IRAS 12302-6317 & 12:33:07.5 & -63:33:41 &  &  & $-0.57\pm{0.14}$ & 1.307 & $11.1^{+5.1}_{-3.3}$ & $13000^{+15000}_{-7000}$ & 1 \\
6058372329633216896 & IRAS 12309-5928 & 12:33:44.5 & -59:45:18 &  &  & $-3.6\pm{1.8}$ & 2.33 & $5.3^{+2.9}_{-1.9}$ & $1000^{+1400}_{-600}$ & 1 \\
\hline
\end{longtable}
\end{landscape}

\begin{landscape}
Table 1 (continued).
\begin{longtable}{llrrlrrrrrr}
\hline
Gaia DR3 source id & SIMBAD name & RA & DEC & SpType & T\textsubscript{eff} & Parallax & RUWE & Distance & Luminosity & Vickers\\
 &  & h:m:s & deg:m:s & & log([K]) & [mas] & & [kpc] & [L$_{\odot}$] & category\\
\hline
6060828565581083264 & IRAS 12360-5740 & 12:38:53.2 & -57:56:31 & F0 & 3.87 & $0.091\pm{0.014}$ & 1.07 & $9.1^{+1.1}_{-0.8}$ & $4500^{+1200}_{-800}$ & 2 \\
6073662099660289536 & IRAS 12419-5414 & 12:44:46.1 & -54:31:12 &  &  & $2.69\pm{0.13}$ & 3.534 & $0.365^{+0.017}_{-0.017}$ & $13.1^{+1.3}_{-1.2}$ & 1 \\
3497154104039422848 & IRAS 12538-2611 & 12:56:30.1 & -26:27:36 & F3Ia & 3.83 & $0.567\pm{0.023}$ & 0.9709 & $1.684^{+0.085}_{-0.065}$ & $6750^{+690}_{-510}$ & 1 \\
6084869868362934144 & IRAS 12584-4837 & 13:01:17.6 & -48:53:20 & B5e & 4.18 & $1.59\pm{0.15}$ & 7.593 & $0.649^{+0.093}_{-0.063}$ & $241^{+74}_{-45}$ & 2 \\
6055883241501706752 & IRAS 13010-6012 & 13:04:05.3 & -60:28:46 &  &  & $-0.02\pm{0.18}$ & 7.212 & $5.5^{+2.5}_{-1.6}$ & $550^{+620}_{-280}$ & 2 \\
5863599857739835264 & IRAS 13064-6103 & 13:09:36.3 & -61:19:37 & B & 4.26 & $-1.36\pm{0.37}$ & 25.79 & $4.5^{+1.8}_{-1.3}$ & $16000^{+15000}_{-8000}$ & 3 \\
6066902993687172608 & IRAS 13110-5425 & 13:14:08.4 & -54:41:34 & F5Ia/ab & 3.81 & $0.64\pm{0.031}$ & 1.06 & $1.535^{+0.071}_{-0.067}$ & $2510^{+240}_{-210}$ & 1 \\
5857811238294426752 & IRAS 13110-6629 & 13:14:26.8 & -66:45:33 &  &  & $0.235\pm{0.021}$ & 0.8677 & $4.00^{+0.30}_{-0.30}$ & $5550^{+860}_{-800}$ & 1 \\
5869845594859880064 & IRAS 13203-5917 & 13:23:32.2 & -59:32:50 & G2I & 3.73 & $0.183\pm{0.042}$ & 3.218 & $5.4^{+1.2}_{-0.7}$ & $2100^{+1100}_{-500}$ & 1 \\
3624198034063995008 & BPS CS22877-0023 & 13:25:39.5 & -08:49:21 & B3 & 4.26 & $0.056\pm{0.028}$ & 1.076 & $9.6^{+4.4}_{-1.8}$ & $820^{+940}_{-280}$ & 1 \\
6083719439934104832 & CD-46 8644 & 13:26:26.2 & -47:16:27 & A7 & 3.9 & $0.177\pm{0.02}$ & 0.933 & $4.89^{+0.44}_{-0.38}$ & $1240^{+240}_{-180}$ & 1 \\
6083708479176016128 & Cl* NGC 5139WOR1957 & 13:27:28.9 & -47:22:48 &  &  & $0.115\pm{0.03}$ & 0.9013 & $6.2^{+1.3}_{-0.8}$ & $1760^{+780}_{-420}$ & 1 \\
6070128028770373888 & IRAS 13245-5036 & 13:27:37.1 & -50:52:05 & A7 & 3.9 & $0.012\pm{0.022}$ & 1.605 & $14.2^{+3.2}_{-2.9}$ & $6900^{+3400}_{-2500}$ & 1 \\
6063703586653222144 & IRAS 13266-5551 & 13:29:51.1 & -56:06:53 & B1Iae & 4.42 & $0.338\pm{0.02}$ & 1.141 & $2.82^{+0.16}_{-0.14}$ & $3560^{+420}_{-360}$ & 1 \\
5783415017323438848 & IRAS 13258-8103 & 13:31:06.3 & -81:18:30 & F8 & 3.78 & $0.72\pm{0.21}$ & 8.569 & $1.61^{+0.70}_{-0.38}$ & $140^{+150}_{-60}$ & 2 \\
5870113880062711552 & IRAS 13313-5838 & 13:34:37.4 & -58:53:33 & K5I & 3.59 & $1.09\pm{0.13}$ & 10.35 & $0.96^{+0.15}_{-0.09}$ & $72^{+24}_{-13}$ & 2 \\
5865398796206273152 & IRAS 13356-6249 & 13:39:06.4 & -63:04:45 &  &  & $0.284\pm{0.093}$ & 1.171 & $2.99^{+0.81}_{-0.64}$ & $5900^{+3600}_{-2200}$ & 1 \\
5869018899564351872 & IRAS 13398-5951 & 13:43:12.4 & -60:07:04 &  &  & $1.53\pm{0.59}$ & 1.628 & $2.3^{+0.8}_{-1.2}$ & $230^{+190}_{-180}$ & 3 \\
5864661779824561664 & IRAS 13416-6243 & 13:45:07.2 & -62:58:15 & G1I & 3.745 & $0.21\pm{0.1}$ & 1.002 & $4.4^{+2.0}_{-1.1}$ & $15000^{+16000}_{-6000}$ & 1 \\
5867561191994328704 & IRAS 13529-5934 & 13:56:24.9 & -59:48:58 &  &  & $-0.91\pm{0.98}$ & 2.641 & $3.9^{+1.6}_{-1.7}$ & $760^{+760}_{-510}$ & 1 \\
5896479309853592448 & IRAS 14072-5446 & 14:10:39.0 & -55:00:27 & A3I & 3.95 & $0.207\pm{0.014}$ & 0.8488 & $4.34^{+0.26}_{-0.24}$ & $4860^{+600}_{-520}$ & 1 \\
5853777267581362176 & IRAS 14103-6311 & 14:14:09.3 & -63:25:47 & M0 & 3.57 & $0.28\pm{0.18}$ & 12.84 & $3.8^{+2.0}_{-1.0}$ & $10000^{+13000}_{-5000}$ & 3 \\
5849962851220246016 & IRAS 14325-6428 & 14:36:34.5 & -64:41:31 & F5I & 3.81 & $0.192\pm{0.037}$ & 2.181 & $4.88^{+0.93}_{-0.62}$ & $4600^{+1900}_{-1100}$ & 2 \\
5849958457496943744 & IRAS 14331-6435 & 14:37:10.3 & -64:48:06 & B3Ie & 4.26 & $0.351\pm{0.058}$ & 3.417 & $2.91^{+0.53}_{-0.44}$ & $4100^{+1600}_{-1100}$ & 1 \\
5877302177877393024 & IRAS 14341-6211 & 14:38:05.1 & -62:24:49 &  &  & $-0.12\pm{0.18}$ & 1.815 & $7.8^{+4.2}_{-2.7}$ & $3500^{+4900}_{-2000}$ & 1 \\
5878583349370477312 & IRAS 14346-5952 & 14:38:24.6 & -60:04:52 &  &  & $0.261\pm{0.096}$ & 1.126 & $3.1^{+1.0}_{-0.7}$ & $4900^{+3500}_{-2000}$ & 2 \\
5906408788891928704 & IRAS 14429-4539 & 14:46:13.7 & -45:52:07 & G0Ie & 3.76 & $-0.11\pm{0.51}$ & 2.788 & $3.8^{+2.7}_{-1.7}$ & $3100^{+5800}_{-2100}$ & 1 \\
5880708980232408448 & IRAS 14482-5725 & 14:51:58.2 & -57:38:17 & A2I & 3.96 & $0.358\pm{0.049}$ & 1.935 & $2.50^{+0.39}_{-0.28}$ & $430^{+140}_{-90}$ & 1 \\
5893945588395282304 & IRAS 14488-5405 & 14:52:28.9 & -54:17:43 & A0Ie & 3.99 & $0.268\pm{0.017}$ & 1.327 & $3.42^{+0.21}_{-0.19}$ & $7400^{+1000}_{-800}$ & 2 \\
5899715786733345536 & IRAS 14562-5406 & 14:59:53.5 & -54:18:09 & W9. & 4.643 & $0.482\pm{0.027}$ & 1.346 & $1.94^{+0.10}_{-0.08}$ & $6830^{+700}_{-540}$ & 3 \\
5903310335089068416 & IRAS 15039-4806 & 15:07:27.2 & -48:17:53 & A5Iab & 3.92 & $0.563\pm{0.026}$ & 0.8877 & $1.708^{+0.074}_{-0.061}$ & $5200^{+460}_{-360}$ & 1 \\
\hline
\end{longtable}
\end{landscape}

\begin{landscape}
Table 1 (continued).
\begin{longtable}{llrrlrrrrrr}
\hline
Gaia DR3 source id & SIMBAD name & RA & DEC & SpType & T\textsubscript{eff} & Parallax & RUWE & Distance & Luminosity & Vickers\\
 &  & h:m:s & deg:m:s & & log([K]) & [mas] & & [kpc] & [L$_{\odot}$] & category\\
\hline
5886573569080505216 & IRAS 15066-5532 & 15:10:26.6 & -55:44:13 &  &  & $0.261\pm{0.025}$ & 1.181 & $3.23^{+0.24}_{-0.20}$ & $2010^{+310}_{-240}$ & 1 \\
5877154701494366336 & IRAS 15103-5754 & 15:14:18.7 & -58:05:20 &  &  & $0.31\pm{0.81}$ & 1.092 & $3.7^{+1.9}_{-1.7}$ & $7200^{+9300}_{-5200}$ & 3 \\
5877096736611221888 & IRAS 15144-5812 & 15:18:21.9 & -58:23:13 &  &  & $-0.018\pm{0.08}$ & 1.06 & $7.3^{+2.1}_{-1.9}$ & $16000^{+10000}_{-7000}$ & 2 \\
5888049732191403904 & IRAS 15154-5258 & 15:19:08.7 & -53:09:51 & W4 & 5.068 & $0.299\pm{0.024}$ & 1.027 & $2.89^{+0.20}_{-0.21}$ & $644^{+93}_{-90}$ & 3 \\
5824126771840487936 & IRAS 15210-6554 & 15:25:31.5 & -66:05:20 & K2I & 3.62 & $-0.15\pm{0.14}$ & 2.279 & $7.5^{+2.6}_{-2.8}$ & $4600^{+3700}_{-2800}$ & 1 \\
1194929381434604800 & IRAS F15240+1452 & 15:26:20.9 & +14:41:36 & B9Iab:p & 4.02 & $0.752\pm{0.079}$ & 2.023 & $1.32^{+0.20}_{-0.13}$ & $1440^{+460}_{-270}$ & 1 \\
1369896865785991424 & BD+33 2642 & 15:51:59.8 & +32:56:54 & O7p & 4.57 & $0.271\pm{0.032}$ & 1.295 & $3.47^{+0.47}_{-0.31}$ & $2260^{+650}_{-380}$ & 2 \\
5996617434442714880 & IRAS 16029-4101 & 16:06:24.4 & -41:09:57 &  &  & $0.2\pm{0.069}$ & 4.272 & $4.3^{+1.0}_{-0.9}$ & $980^{+480}_{-360}$ & 1 \\
5933063252888145920 & IRAS 16086-5255 & 16:12:30.6 & -53:03:11 &  &  & $0.358\pm{0.014}$ & 0.912 & $2.50^{+0.12}_{-0.10}$ & $771^{+79}_{-58}$ & 2 \\
5835411094089870592 & IRAS 16099-5651 & 16:14:01.4 & -56:59:28 & M8 & 3.47 & $0.34\pm{0.11}$ & 0.8452 & $3.9^{+2.0}_{-1.0}$ & $12000^{+16000}_{-5000}$ & 3 \\
5935061172876722176 & IRAS 16115-5044 & 16:15:17.9 & -50:52:21 &  &  & $0.291\pm{0.09}$ & 0.9794 & $3.05^{+0.80}_{-0.80}$ & $13100^{+7900}_{-6000}$ & 1 \\
5990014935825542144 & IRAS 16130-4620 & 16:16:42.9 & -46:27:53 &  &  & $-0.89\pm{0.54}$ & 1.115 & $5.4^{+3.5}_{-1.9}$ & $2900^{+5000}_{-1700}$ & 2 \\
5934701559547878144 & IRAS 16133-5151 & 16:17:13.2 & -51:59:11 & B0 & 4.49 & $0.366\pm{0.086}$ & 2.51 & $2.65^{+0.62}_{-0.46}$ & $17400^{+9100}_{-5500}$ & 3 \\
5932016212933920384 & LS 3593 & 16:24:39.4 & -54:38:08 & A0Ib & 3.99 & $0.443\pm{0.019}$ & 0.8017 & $2.138^{+0.091}_{-0.080}$ & $4450^{+390}_{-330}$ & 1 \\
5831295999979910656 & IRAS 16206-5956 & 16:25:02.9 & -60:03:32 & A0Iae & 3.99 & $0.182\pm{0.019}$ & 1.066 & $5.02^{+0.66}_{-0.41}$ & $6000^{+1700}_{-1000}$ & 1 \\
4351018375858237952 & IRAS F16277-0724 & 16:30:30.1 & -07:30:52 & A7Ib & 3.9 & $0.281\pm{0.025}$ & 1.156 & $3.10^{+0.29}_{-0.21}$ & $40700^{+7900}_{-5200}$ & 2 \\
5941189713256986752 & IRAS 16279-4757 & 16:31:38.8 & -48:04:07 & M3II & 3.49 & $-0.08\pm{0.19}$ & 1.465 & $4.5^{+2.0}_{-1.3}$ & $28000^{+31000}_{-14000}$ & 1 \\
5943899425344608000 & IRAS 16283-4424 & 16:31:57.6 & -44:31:25 &  &  & $-0.095\pm{0.039}$ & 2.773 & $13.8^{+3.2}_{-2.9}$ & $11100^{+5700}_{-4100}$ & 1 \\
1324742534573959424 & BD+32 2754 & 16:36:11.9 & +32:29:21 & F8 & 3.78 & $3.239\pm{0.014}$ & 1.161 & $0.3067^{+0.0012}_{-0.0012}$ & $10.340^{+0.090}_{-0.080}$ & 2 \\
5943047784868946688 & IRAS 16328-4517 & 16:36:25.8 & -45:24:04 &  &  & $0.64\pm{0.2}$ & 6.352 & $1.94^{+0.73}_{-0.52}$ & $310^{+280}_{-140}$ & 1 \\
5941041038676090112 & IRAS 16333-4807 & 16:37:06.8 & -48:13:43 &  &  & $0.364\pm{0.024}$ & 1.003 & $2.43^{+0.14}_{-0.11}$ & $1750^{+210}_{-160}$ & 3 \\
1328057763997734144 & Cl* NGC 6205BARN29 & 16:41:33.8 & +36:26:11 & B2p & 4.36 & $0.077\pm{0.03}$ & 1.157 & $9.0^{+2.7}_{-1.9}$ & $2200^{+1500}_{-800}$ & 1 \\
4334241408966611328 & IRAS 16476-1122 & 16:50:24.3 & -11:27:58 & M1I & 3.56 & $0.471\pm{0.031}$ & 1.099 & $1.98^{+0.13}_{-0.12}$ & $1010^{+140}_{-120}$ & 2 \\
5969973999973524224 & IRAS 16494-3930 & 16:52:55.4 & -39:34:58 &  &  & $0.33\pm{0.059}$ & 1.014 & $2.92^{+0.56}_{-0.41}$ & $590^{+250}_{-150}$ & 1 \\
4365451214021224320 & LS IV-04$\_$1 & 16:56:27.8 & -04:47:24 & B & 4.26 & $0.033\pm{0.017}$ & 1.03 & $10.9^{+1.5}_{-1.4}$ & $4200^{+1300}_{-1000}$ & 1 \\
6029425384023553792 & IRAS 16559-2957 & 16:59:08.2 & -30:01:41 & F5I(e) & 3.81 & $-2.03\pm{0.5}$ & 23.13 & $7.2^{+2.7}_{-1.9}$ & $10400^{+9100}_{-4700}$ & 1 \\
5963059480546004608 & IRAS 16594-4656 & 17:03:10.0 & -47:00:29 & B7 & 4.1 & $0.546\pm{0.073}$ & 0.6971 & $1.72^{+0.24}_{-0.19}$ & $2600^{+760}_{-550}$ & 1 \\
5965575811694758784 & IRAS 17009-4154 & 17:04:29.8 & -41:58:38 &  &  & $-2.73\pm{0.86}$ & 1.26 & $6.0^{+4.5}_{-2.8}$ & $18000^{+37000}_{-13000}$ & 1 \\
4568163710366782848 & PG 1704+222 & 17:06:46.2 & +22:05:51 & B3 & 4.26 & $0.091\pm{0.025}$ & 0.9389 & $7.5^{+1.8}_{-1.2}$ & $1240^{+690}_{-370}$ & 1 \\
5917195002285834240 & IRAS 17047-5650 & 17:09:00.8 & -56:54:48 & W9 & 4.643 & $0.678\pm{0.047}$ & 2.916 & $1.48^{+0.11}_{-0.10}$ & $5580^{+850}_{-700}$ & 3 \\
5973374200987325056 & IRAS 17067-3759 & 17:10:08.3 & -38:03:23 &  &  & $-0.1\pm{0.31}$ & 0.9856 & $4.5^{+2.2}_{-1.6}$ & $1100^{+1300}_{-600}$ & 2 \\
\hline
\end{longtable}
\end{landscape}

\begin{landscape}
Table 1 (continued).
\begin{longtable}{llrrlrrrrrr}
\hline
Gaia DR3 source id & SIMBAD name & RA & DEC & SpType & T\textsubscript{eff} & Parallax & RUWE & Distance & Luminosity & Vickers\\
 &  & h:m:s & deg:m:s & & log([K]) & [mas] & & [kpc] & [L$_{\odot}$] & category\\
\hline
4128590918794710272 & IRAS 17074-1845 & 17:10:24.0 & -18:49:01 & B5Ibe & 4.18 & $0.072\pm{0.042}$ & 1.707 & $8.0^{+1.9}_{-1.6}$ & $8200^{+4200}_{-2900}$ & 1 \\
4128590918794710272 & IRAS 17074-1845 & 17:10:24.0 & -18:49:01 & B5Ibe & 4.18 & $0.072\pm{0.042}$ & 1.707 & $8.0^{+1.9}_{-1.6}$ & $6700^{+3500}_{-2400}$ & 3 \\
5980714063986945920 & IRAS 17106-3046 & 17:13:51.8 & -30:49:42 & F5I & 3.81 & $-0.06\pm{0.16}$ & 4.161 & $7.2^{+2.3}_{-2.0}$ & $12800^{+9400}_{-6200}$ & 1 \\
5971985036767008256 & IRAS 17130-4029 & 17:16:28.9 & -40:32:30 &  &  & $-0.24\pm{0.48}$ & 1.961 & $5.6^{+2.9}_{-2.1}$ & $1200^{+1500}_{-700}$ & 2 \\
5972460442434579968 & IRAS 17163-3907 & 17:19:49.3 & -39:10:37 &  &  & $0.126\pm{0.065}$ & 0.9698 & $5.2^{+1.4}_{-1.0}$ & $580000^{+350000}_{-210000}$ & 2 \\
4058987022459300096 & IRAS 17183-3017 & 17:21:31.9 & -30:20:50 &  &  & $0.01\pm{0.23}$ & 3.618 & $7.2^{+2.2}_{-2.1}$ & $6700^{+4600}_{-3300}$ & 3 \\
4108072721831278976 & IRAS 17195-2710 & 17:22:43.6 & -27:13:38 &  &  & $0.12\pm{0.4}$ & 2.942 & $6.9^{+2.5}_{-2.8}$ & $8100^{+7000}_{-5300}$ & 1 \\
4136944866387751552 & IRAS 17203-1534 & 17:23:11.9 & -15:37:16 & B1IIIpe & 4.42 & $0.066\pm{0.025}$ & 1.379 & $7.9^{+1.3}_{-1.2}$ & $9200^{+3300}_{-2500}$ & 1 \\
4109870908774509184 & IRAS 17209-2556A & 17:24:01.4 & -25:59:23 & W9 & 4.643 & $0.17\pm{0.032}$ & 1.189 & $5.10^{+0.93}_{-0.71}$ & $1200^{+480}_{-310}$ & 3 \\
5972489407656030720 & IRAS 17208-3859 & 17:24:19.5 & -39:01:48 & A2I & 3.96 & $0.68\pm{0.49}$ & 0.9974 & $5.0^{+2.7}_{-2.6}$ & $3200^{+4300}_{-2400}$ & 1 \\
4109553493474085504 & IRAS 17223-2659 & 17:25:26.5 & -27:02:01 &  &  & $0.163\pm{0.057}$ & 1.747 & $5.2^{+1.5}_{-1.2}$ & $4200^{+2800}_{-1700}$ & 2 \\
5960186761599871488 & IRAS 17234-4008 & 17:26:56.1 & -40:11:02 &  &  & $0.2\pm{0.1}$ & 1.063 & $5.4^{+3.5}_{-1.5}$ & $1600^{+2700}_{-800}$ & 1 \\
5975007147549090432 & IRAS 17251-3505 & 17:28:27.7 & -35:07:33 &  &  & $0.59\pm{0.23}$ & 0.9569 & $2.4^{+1.1}_{-0.8}$ & $1900^{+2200}_{-1000}$ & 3 \\
4162959693758887424 & IRAS 17279-1119 & 17:30:46.8 & -11:22:10 & F2/3II & 3.85 & $0.201\pm{0.016}$ & 1.022 & $4.27^{+0.31}_{-0.29}$ & $9300^{+1400}_{-1200}$ & 1 \\
5975083327400970752 & IRAS 17277-3506 & 17:31:04.1 & -35:08:41 & B4 & 4.22 & $0.743\pm{0.025}$ & 1.111 & $1.285^{+0.051}_{-0.042}$ & $6800^{+550}_{-440}$ & 2 \\
4110550475656804864 & IRAS 17291-2402 & 17:32:12.8 & -24:04:59 &  &  & $0.74\pm{0.4}$ & 10.87 & $3.8^{+2.1}_{-1.8}$ & $1900^{+2600}_{-1400}$ & 1 \\
4061265519768800768 & IRAS 17317-2743 & 17:34:53.6 & -27:45:13 & F5I & 3.81 & $-0.06\pm{0.12}$ & 2.93 & $8.3^{+2.4}_{-2.2}$ & $8700^{+5700}_{-4000}$ & 1 \\
5946845601071213696 & IRAS 17311-4924 & 17:35:02.4 & -49:26:27 & B1Iae & 4.42 & $0.238\pm{0.02}$ & 1.084 & $3.86^{+0.29}_{-0.21}$ & $10000^{+1500}_{-1100}$ & 1 \\
4117592465329067008 & IRAS 17332-2215 & 17:36:17.0 & -22:17:20 & K2I & 3.62 & $3.2\pm{0.57}$ & 3.549 & $0.35^{+0.10}_{-0.06}$ & $9.7^{+6.6}_{-3.2}$ & 1 \\
5955201232284272384 & [DSH2001] 279-19 & 17:39:02.3 & -45:00:38 &  &  & $0.174\pm{0.022}$ & 0.8732 & $4.71^{+0.58}_{-0.41}$ & $204^{+53}_{-34}$ & 2 \\
4060159376692627840 & IRAS 17358-2854 & 17:39:02.9 & -28:56:37 &  &  & $0.381\pm{0.034}$ & 0.9727 & $2.28^{+0.15}_{-0.16}$ & $626^{+84}_{-86}$ & 3 \\
4117263131638476032 & IRAS 17360-2142 & 17:39:05.8 & -21:43:52 &  &  & $-0.21\pm{0.19}$ & 4.72 & $7.2^{+1.8}_{-2.0}$ & $3500^{+2000}_{-1700}$ & 1 \\
4053492968991830400 & IRAS 17370-3357 & 17:40:20.1 & -33:59:15 &  &  & $0.47\pm{0.68}$ & 1.098 & $8.1^{+2.8}_{-3.4}$ & $4300^{+3600}_{-2900}$ & 1 \\
4118178784208141568 & IRAS 17376-2040 & 17:40:38.7 & -20:41:52 &  &  & $-3.3\pm{1.1}$ & 1.036 & $7.0^{+2.2}_{-1.9}$ & $6800^{+4800}_{-3100}$ & 2 \\
4124125282361429504 & IRAS 17381-1616 & 17:41:00.1 & -16:18:13 & B1Ibe & 4.42 & $0.109\pm{0.025}$ & 1.148 & $6.4^{+1.1}_{-0.8}$ & $3000^{+1100}_{-700}$ & 1 \\
4053880649927702656 & IRAS 17385-3332 & 17:41:52.4 & -33:33:41 &  &  & $-0.25\pm{0.16}$ & 3.945 & $9.1^{+3.2}_{-2.9}$ & $5400^{+4500}_{-2900}$ & 1 \\
1367102319545324288 & IRAS 17436+5003 & 17:44:55.3 & +50:02:39 & F3Ib & 3.83 & $0.502\pm{0.024}$ & 1.216 & $1.92^{+0.10}_{-0.09}$ & $6980^{+700}_{-650}$ & 1 \\
4120637086125583360 & IRAS 17423-1755 & 17:45:14.1 & -17:56:47 & B7e & 4.1 & $0.31\pm{0.3}$ & 1.665 & $5.3^{+2.2}_{-1.4}$ & $11000^{+11000}_{-5000}$ & 1 \\
4120632688077368192 & IRAS 17433-1750 & 17:46:15.7 & -17:51:47 &  &  & $0.032\pm{0.02}$ & 0.9203 & $9.6^{+1.7}_{-1.3}$ & $16400^{+6200}_{-4200}$ & 2 \\
4041945343757244160 & IRAS 17440-3310 & 17:47:22.6 & -33:11:08 & F3I & 3.83 & $-0.16\pm{0.12}$ & 0.7693 & $8.3^{+2.4}_{-2.0}$ & $8000^{+5200}_{-3400}$ & 1 \\
5954670408684703872 & IRAS 17476-4446 & 17:51:16.5 & -44:47:30 & B7Ie & 4.1 & $0.128\pm{0.027}$ & 1.285 & $6.2^{+1.9}_{-0.9}$ & $1030^{+730}_{-270}$ & 1 \\
\hline
\end{longtable}
\end{landscape}

\begin{landscape}
Table 1 (continued).
\begin{longtable}{llrrlrrrrrr}
\hline
Gaia DR3 source id & SIMBAD name & RA & DEC & SpType & T\textsubscript{eff} & Parallax & RUWE & Distance & Luminosity & Vickers\\
 &  & h:m:s & deg:m:s & & log([K]) & [mas] & & [kpc] & [L$_{\odot}$] & category\\
\hline
4056355822397882880 & IRAS 17480-3023 & 17:51:19.0 & -30:23:52 & W9 & 4.643 & $0.083\pm{0.062}$ & 0.9587 & $8.2^{+3.3}_{-2.0}$ & $4500^{+4400}_{-2000}$ & 3 \\
4119492494479957760 & IRAS 17487-1922 & 17:51:45.0 & -19:23:45 &  &  & $-0.84\pm{0.49}$ & 3.084 & $5.9^{+3.1}_{-2.2}$ & $3500^{+4600}_{-2100}$ & 1 \\
4067343654354938112 & IRAS 17516-2525 & 17:54:43.3 & -25:26:29 & B0 & 4.49 & $0.19\pm{0.073}$ & 0.9617 & $4.3^{+1.4}_{-1.2}$ & $28000^{+22000}_{-13000}$ & 1 \\
4582795323914832000 & IRAS 17534+2603 & 17:55:25.1 & +26:03:00 & F2Ibp & 3.85 & $0.761\pm{0.061}$ & 1.223 & $1.31^{+0.13}_{-0.11}$ & $15300^{+3200}_{-2400}$ & 1 \\
4172337943816530432 & IRAS 17542-0603 & 17:56:56.0 & -06:04:10 &  &  & $0.1\pm{0.086}$ & 5.976 & $5.9^{+2.1}_{-1.3}$ & $4700^{+3900}_{-1900}$ & 1 \\
4035783611879028736 & IRAS 17567-3849 & 17:57:47.7 & -38:34:12 &  &  & $-0.21\pm{0.21}$ & 3.581 & $7.2^{+3.4}_{-2.2}$ & $6900^{+8200}_{-3500}$ & 3 \\
4062759511236966016 & IRAS 17550-2800 & 17:58:10.4 & -28:00:31 &  &  & $-0.08\pm{0.21}$ & 1.578 & $7.1^{+3.5}_{-2.1}$ & $3500^{+4300}_{-1800}$ & 2 \\
4062301564840251520 & IRAS 17574-2921 & 18:00:37.6 & -29:21:50 & W9 & 4.643 & $0.091\pm{0.052}$ & 1.03 & $7.7^{+2.1}_{-1.9}$ & $1800^{+1200}_{-800}$ & 3 \\
4062817927079802240 & IRAS 18022-2822 & 18:05:25.7 & -28:22:03 & W9 & 4.643 & $0.237\pm{0.043}$ & 1.073 & $3.92^{+0.82}_{-0.66}$ & $740^{+340}_{-230}$ & 3 \\
4042544062195042176 & IRAS 18023-3409 & 18:05:38.4 & -34:09:31 & B9Ia+e & 4.02 & $0.022\pm{0.027}$ & 1.138 & $9.4^{+1.1}_{-1.2}$ & $23200^{+6000}_{-5700}$ & 1 \\
4035907203854415488 & IRAS 18025-3906 & 18:06:03.3 & -39:05:56 & G2I & 3.73 & $0.54\pm{0.19}$ & 8.582 & $3.0^{+4.1}_{-1.1}$ & $2100^{+9400}_{-1200}$ & 1 \\
4579182637944779264 & IRAS 18062+2410 & 18:08:19.9 & +24:10:44 & B1IIIe & 4.42 & $0.143\pm{0.049}$ & 2.886 & $4.5^{+1.1}_{-0.9}$ & $3600^{+2000}_{-1300}$ & 1 \\
4065774307705264384 & IRAS 18061-2505 & 18:09:12.4 & -25:04:33 & W8 & 4.778 & $0.55\pm{0.12}$ & 0.9664 & $2.18^{+0.71}_{-0.60}$ & $430^{+330}_{-200}$ & 3 \\
4158154754919296000 & IRAS 18075-0924 & 18:10:15.1 & -09:23:34 & G2I & 3.73 & $-0.17\pm{0.19}$ & 7.58 & $6.9^{+2.8}_{-2.5}$ & $9700^{+9200}_{-5700}$ & 1 \\
4093773847992815744 & IRAS 18083-2155 & 18:11:19.8 & -21:54:59 &  &  & $-0.51\pm{0.4}$ & 0.9837 & $6.7^{+3.7}_{-2.7}$ & $7000^{+10000}_{-5000}$ & 1 \\
4580154606223711872 & IRAS 18095+2704 & 18:11:30.5 & +27:05:15 & F3Ib & 3.83 & $0.03\pm{0.18}$ & 13.74 & $4.9^{+1.7}_{-1.6}$ & $17000^{+13000}_{-9000}$ & 1 \\
4049331244394134912 & IRAS 18129-3053 & 18:13:58.6 & -30:36:57 & W9 & 4.643 & $0.34\pm{0.11}$ & 3.771 & $4.0^{+2.3}_{-1.2}$ & $7000^{+11000}_{-4000}$ & 3 \\
4065347387968755328 & IRAS 18113-2503 & 18:14:27.2 & -25:03:01 &  &  & $0.144\pm{0.035}$ & 1.123 & $5.2^{+1.0}_{-0.8}$ & $2010^{+810}_{-550}$ & 1 \\
4049379725984965120 & LS 4825 & 18:16:00.5 & -30:45:24 & B1Ib & 4.42 & $0.003\pm{0.018}$ & 0.4916 & $10.0^{+1.3}_{-0.8}$ & $9500^{+2800}_{-1500}$ & 2 \\
4065322923848085760 & IRAS 18170-2416 & 18:20:08.8 & -24:15:04 & W9 & 4.643 & $0.033\pm{0.032}$ & 0.9947 & $8.6^{+2.1}_{-1.2}$ & $3300^{+1800}_{-900}$ & 3 \\
4046476534251259904 & CD-30 15602 & 18:22:42.6 & -30:14:40 & G0: & 3.76 & $0.118\pm{0.025}$ & 0.6698 & $5.59^{+0.93}_{-0.71}$ & $1220^{+440}_{-290}$ & 2 \\
4270080782317422208 & IRAS 18240-0244 & 18:26:40.2 & -02:42:57 & W7 & 4.851 & $0.15\pm{0.22}$ & 0.7685 & $4.1^{+1.5}_{-1.7}$ & $5000^{+4300}_{-3300}$ & 3 \\
4155847571502019840 & IRAS 18286-0959 & 18:31:22.9 & -09:57:20 &  &  & $0.5\pm{0.58}$ & 1.027 & $3.3^{+2.6}_{-1.4}$ & $7000^{+15000}_{-5000}$ & 1 \\
4093689116883710592 & IRAS 18313-1738 & 18:34:16.2 & -17:36:15 & B & 4.26 & $1.244\pm{0.051}$ & 2.397 & $0.793^{+0.033}_{-0.030}$ & $148^{+13}_{-11}$ & 2 \\
4104509518289438720 & IRAS 18321-1401 & 18:34:57.6 & -13:58:49 &  &  & $2.7\pm{1.3}$ & 0.9067 & $5.5^{+3.1}_{-2.5}$ & $520^{+740}_{-360}$ & 1 \\
2096072103492979584 & V534 Lyr & 18:37:58.9 & +37:26:06 & A0Iabe & 3.99 & $0.538\pm{0.025}$ & 1.081 & $1.807^{+0.084}_{-0.081}$ & $1930^{+180}_{-170}$ & 2 \\
6736747708089687936 & IRAS 18371-3159 & 18:40:21.9 & -31:56:49 & B1Iabe & 4.42 & $0.085\pm{0.024}$ & 0.9848 & $7.3^{+1.3}_{-0.9}$ & $3200^{+1300}_{-700}$ & 1 \\
4099619470274753408 & IRAS 18379-1707 & 18:40:48.7 & -17:04:39 & B1IIIep & 4.42 & $0.142\pm{0.023}$ & 0.9967 & $5.35^{+0.69}_{-0.55}$ & $3270^{+900}_{-640}$ & 1 \\
4508727788268978176 & IRAS 18385+1350 & 18:40:51.8 & +13:52:51 &  &  & $0.029\pm{0.08}$ & 0.9762 & $7.4^{+2.5}_{-1.9}$ & $1040^{+840}_{-470}$ & 1 \\
4072427555640528000 & IRAS 18384-2800 & 18:41:37.0 & -27:57:01 & F2/3Ia & 3.85 & $0.11\pm{0.027}$ & 1.346 & $7.0^{+1.3}_{-1.2}$ & $37000^{+15000}_{-12000}$ & 1 \\
4253622708258596736 & IRAS 18420-0512 & 18:44:41.7 & -05:09:15 &  &  & $-0.33\pm{0.21}$ & 1.105 & $7.6^{+2.6}_{-2.1}$ & $7200^{+6000}_{-3500}$ & 2 \\
\hline
\end{longtable}
\end{landscape}

\begin{landscape}
Table 1 (continued).
\begin{longtable}{llrrlrrrrrr}
\hline
Gaia DR3 source id & SIMBAD name & RA & DEC & SpType & T\textsubscript{eff} & Parallax & RUWE & Distance & Luminosity & Vickers\\
 &  & h:m:s & deg:m:s & & log([K]) & [mas] & & [kpc] & [L$_{\odot}$] & category\\
\hline
4266422531037747200 & IRAS 18454+0001 & 18:48:01.2 & +00:04:50 &  &  & $0.63\pm{0.71}$ & 1.204 & $4.6^{+2.4}_{-2.2}$ & $1200^{+1600}_{-900}$ & 2 \\
4285847607265349632 & IRAS 18485+0642 & 18:50:59.7 & +06:46:00 &  &  & $0.5\pm{0.51}$ & 1.706 & $3.5^{+1.7}_{-1.5}$ & $1400^{+1700}_{-1000}$ & 1 \\
4252449701157768960 & IRAS 18489-0629 & 18:51:39.2 & -06:26:06 &  &  & $-0.018\pm{0.056}$ & 2.337 & $10.1^{+2.7}_{-1.9}$ & $41000^{+24000}_{-14000}$ & 1 \\
4203848980711226112 & Cl* NGC6712SSC26 & 18:53:05.7 & -08:42:36 &  &  & $0.089\pm{0.016}$ & 0.7811 & $7.63^{+0.62}_{-0.71}$ & $700^{+120}_{-120}$ & 2 \\
4282499452616310912 & IRAS 18539+0549 & 18:56:22.7 & +05:53:00 &  &  & $0.906\pm{0.023}$ & 1.1 & $1.083^{+0.031}_{-0.027}$ & $39.8^{+2.3}_{-1.9}$ & 2 \\
4265817151122002688 & IRAS 18582+0001 & 19:00:49.0 & +00:06:14 &  &  & $0.0\pm{0.47}$ & 1.057 & $5.8^{+3.2}_{-2.4}$ & $5100^{+7200}_{-3400}$ & 1 \\
4080575933190651008 & IRAS 19016-2330 & 19:04:43.5 & -23:26:10 &  &  & $1.08\pm{0.28}$ & 6.334 & $1.02^{+0.43}_{-0.21}$ & $290^{+290}_{-110}$ & 3 \\
6715619076008049792 & LSE 148 & 19:07:07.9 & -41:43:18 & B5 & 4.18 & $0.859\pm{0.049}$ & 1.093 & $1.130^{+0.060}_{-0.053}$ & $1250^{+140}_{-120}$ & 1 \\
4293369057089082112 & IRAS 19075+0432 & 19:09:59.8 & +04:37:09 &  &  & $0.161\pm{0.027}$ & 1.119 & $4.90^{+0.79}_{-0.60}$ & $5300^{+1800}_{-1200}$ & 2 \\
4264026012336768000 & IRAS 19114+0002 & 19:13:58.7 & +00:07:32 & G2Ia & 3.73 & $0.189\pm{0.021}$ & 0.9219 & $4.43^{+0.35}_{-0.35}$ & $181000^{+30000}_{-28000}$ & 1 \\
4520072476226779520 & IRAS 19134+2131 & 19:15:35.1 & +21:36:34 &  &  & $0.13\pm{0.82}$ & 1.601 & $5.8^{+2.9}_{-2.5}$ & $2800^{+3500}_{-1900}$ & 1 \\
6435349718091211264 & LSE 237 & 19:18:49.1 & -64:35:32 & B5 & 4.18 & $0.069\pm{0.038}$ & 0.8585 & $7.7^{+2.0}_{-1.5}$ & $1550^{+910}_{-540}$ & 1 \\
4515084168071571840 & IRAS 19181+1806 & 19:20:24.8 & +18:11:43 &  &  & $0.47\pm{0.56}$ & 2.219 & $5.2^{+2.9}_{-3.0}$ & $700^{+1000}_{-600}$ & 2 \\
2049984454412871296 & IRAS 19200+3457 & 19:21:55.3 & +35:02:57 & B8 & 4.06 & $0.025\pm{0.019}$ & 1.066 & $10.3^{+2.0}_{-1.3}$ & $7200^{+3100}_{-1700}$ & 2 \\
4516723883521069952 & IRAS 19207+2023 & 19:22:55.9 & +20:28:55 & F6I & 3.8 & $-0.41\pm{0.13}$ & 8.057 & $9.9^{+3.8}_{-2.7}$ & $5700^{+5200}_{-2700}$ & 1 \\
2018131400704379648 & IRAS 19255+2123 & 19:27:43.9 & +21:30:04 &  &  & $0.17\pm{0.11}$ & 1.004 & $4.9^{+2.3}_{-1.5}$ & $4300^{+4900}_{-2300}$ & 3 \\
6744366945682210560 & IRAS 19288-3419 & 19:32:06.8 & -34:12:59 & W4 & 5.068 & $-0.52\pm{0.46}$ & 13.91 & $5.8^{+3.9}_{-2.0}$ & $190^{+360}_{-110}$ & 3 \\
4318134628803970816 & IRAS 19306+1407 & 19:32:54.9 & +14:13:36 & B0-1I & 4.49 & $0.085\pm{0.059}$ & 1.688 & $7.1^{+2.1}_{-1.7}$ & $8500^{+5600}_{-3600}$ & 1 \\
2025220089635846528 & IRAS 19309+2646 & 19:32:57.6 & +26:52:45 &  &  & $0.18\pm{0.13}$ & 1.869 & $5.0^{+2.3}_{-1.3}$ & $2500^{+2900}_{-1200}$ & 3 \\
1825500124644105728 & IRAS 19312+1950 & 19:33:24.2 & +19:56:57 &  &  & $0.42\pm{0.24}$ & 1.062 & $2.8^{+1.5}_{-0.7}$ & $8000^{+11000}_{-4000}$ & 2 \\
2032744769234150016 & IRAS 19327+3024 & 19:34:45.3 & +30:30:57 & W9 & 4.643 & $0.618\pm{0.032}$ & 0.9284 & $1.562^{+0.068}_{-0.071}$ & $6210^{+560}_{-550}$ & 3 \\
2032364166432389760 & IRAS 19343+2926 & 19:36:18.8 & +29:32:49 & B0 & 4.49 & $0.628\pm{0.093}$ & 1.606 & $1.57^{+0.23}_{-0.19}$ & $2160^{+680}_{-480}$ & 1 \\
4301238979044890496 & IRAS 19356+0754 & 19:38:01.2 & +08:01:32 &  &  & $0.318\pm{0.094}$ & 2.418 & $3.3^{+1.1}_{-0.8}$ & $780^{+630}_{-340}$ & 1 \\
2020388045260203008 & IRAS 19374+2359 & 19:39:35.7 & +24:06:28 & B4 & 4.22 & $-1.16\pm{0.47}$ & 1.873 & $5.2^{+2.4}_{-2.1}$ & $15000^{+17000}_{-10000}$ & 1 \\
2022052808961769088 & IRAS 19376+2622 & 19:39:43.3 & +26:29:35 &  &  & $0.5\pm{0.15}$ & 2.686 & $2.5^{+1.1}_{-0.8}$ & $2900^{+3100}_{-1600}$ & 3 \\
4240112390324832384 & IRAS 19386+0155 & 19:41:08.3 & +02:02:30 & F5Ib & 3.81 & $0.32\pm{0.16}$ & 11.59 & $3.6^{+2.0}_{-1.2}$ & $7000^{+10000}_{-4000}$ & 1 \\
4318934003785783680 & IRAS 19396+1637 & 19:41:57.0 & +16:44:39 & M7 & 3.47 & $0.97\pm{0.1}$ & 5.74 & $1.04^{+0.12}_{-0.10}$ & $1930^{+470}_{-360}$ & 3 \\
2049034819957965312 & IRAS 19410+3733 & 19:42:53.0 & +37:40:42 & F3Ib & 3.83 & $1.042\pm{0.02}$ & 1.082 & $0.945^{+0.019}_{-0.024}$ & $1016^{+41}_{-51}$ & 2 \\
2031794791233840128 & IRAS 19454+2920 & 19:47:24.8 & +29:28:11 & A0 & 3.99 & $0.098\pm{0.025}$ & 1.048 & $6.8^{+1.3}_{-0.9}$ & $18300^{+7800}_{-4700}$ & 1 \\
2033763428091006720 & IRAS 19475+3119 & 19:49:29.4 & +31:27:16 & F3Ibe & 3.83 & $0.316\pm{0.021}$ & 1.429 & $2.97^{+0.16}_{-0.18}$ & $22300^{+2500}_{-2700}$ & 1 \\
2020571869841643392 & IRAS 19477+2401 & 19:49:55.0 & +24:08:55 & F5I & 3.81 & $0.56\pm{0.62}$ & 2.321 & $4.1^{+2.1}_{-1.9}$ & $4300^{+5400}_{-3100}$ & 1 \\
\hline
\end{longtable}
\end{landscape}

\renewcommand{\arraystretch}{1.2}
\begin{landscape}
Table 1 (continued).
\begin{longtable}{llrrlrrrrrr}
\hline
Gaia DR3 source id & SIMBAD name & RA & DEC & SpType & T\textsubscript{eff} & Parallax & RUWE & Distance & Luminosity & Vickers\\
 &  & h:m:s & deg:m:s & & log([K]) & [mas] & & [kpc] & [L$_{\odot}$] & category\\
\hline
2026709201998745472 & IRAS 19480+2504 & 19:50:08.3 & +25:11:59 & C-rich & & $0.7\pm{0.34}$ & 1.006 & $3.4^{+2.3}_{-1.7}$ & $4700^{+8500}_{-3600}$ & 1 \\
6871175064823382912 & IRAS 19500-1709 & 19:52:52.7 & -17:01:49 & F4Ia & 3.82 & $0.399\pm{0.031}$ & 1.009 & $2.31^{+0.17}_{-0.14}$ & $5460^{+840}_{-660}$ & 1 \\
2074302426124470656 & IRAS 19589+4020 & 20:00:42.9 & +40:29:09 & F5I & 3.81 & $-0.47\pm{0.16}$ & 10.96 & $10.0^{+4.4}_{-3.0}$ & $5200^{+5600}_{-2700}$ & 1 \\
6879196723703009920 & IRAS 19590-1249 & 20:01:49.9 & -12:41:17 & B1Ibe & 4.42 & $0.152\pm{0.026}$ & 0.9832 & $4.73^{+0.68}_{-0.58}$ & $1750^{+540}_{-400}$ & 1 \\
2034134414507432064 & IRAS 20000+3239 & 20:01:59.5 & +32:47:32 & G2I & 3.73 & $0.205\pm{0.049}$ & 2.246 & $4.6^{+1.5}_{-0.9}$ & $10300^{+7800}_{-3600}$ & 1 \\
2030200671149815424 & IRAS 20004+2955 & 20:02:27.2 & +30:04:25 & G7Ia & 3.67 & $0.239\pm{0.018}$ & 0.9075 & $3.72^{+0.26}_{-0.24}$ & $52200^{+7700}_{-6400}$ & 1 \\
4190636669164572928 & IRAS 20023-1144 & 20:05:05.3 & -11:35:58 & F2II & 3.85 & $1.336\pm{0.027}$ & 0.8088 & $0.727^{+0.016}_{-0.013}$ & $1763^{+76}_{-64}$ & 2 \\
2055174630333744384 & IRAS 20042+3259 & 20:06:10.6 & +33:07:51 &  &  & $0.409\pm{0.037}$ & 1.179 & $2.29^{+0.20}_{-0.19}$ & $110^{+20}_{-17}$ & 1 \\
2060806470651334912 & IRAS 20094+3721 & 20:11:16.6 & +37:30:52 &  &  & $0.53\pm{0.011}$ & 1.029 & $1.787^{+0.038}_{-0.041}$ & $1412^{+61}_{-64}$ & 2 \\
1803364717856260736 & IRAS 20136+1309 & 20:16:00.6 & +13:18:56 & F5I & 3.81 & $0.279\pm{0.074}$ & 4.92 & $3.7^{+1.6}_{-0.8}$ & $1000^{+1000}_{-400}$ & 1 \\
2060616220769708672 & IRAS 20145+3656 & 20:16:26.3 & +37:06:09 & B & 4.26 & $-0.15\pm{0.074}$ & 3.182 & $9.4^{+2.6}_{-1.8}$ & $60000^{+37000}_{-20000}$ & 3 \\
1836195688380634368 & IRAS 20160+2734 & 20:18:05.9 & +27:44:04 & F3Ie & 3.83 & $0.391\pm{0.016}$ & 1.041 & $2.317^{+0.093}_{-0.080}$ & $4080^{+330}_{-280}$ & 1 \\
2054521833963867008 & IRAS 20174+3222 & 20:19:27.9 & +32:32:16 &  &  & $0.124\pm{0.044}$ & 1.365 & $5.2^{+1.4}_{-0.9}$ & $2100^{+1200}_{-600}$ & 1 \\
2056435602670418688 & IRAS 20244+3509 & 20:26:25.0 & +35:19:13 &  &  & $0.528\pm{0.026}$ & 1.144 & $1.694^{+0.074}_{-0.063}$ & $299^{+26}_{-22}$ & 2 \\
1859190569633451648 & IRAS 20406+2953 & 20:42:45.9 & +30:04:08 &  &  & $0.21\pm{0.12}$ & 2.219 & $3.9^{+1.4}_{-0.9}$ & $4600^{+3900}_{-1900}$ & 3 \\
1869422453048750336 & IRAS 20462+3416 & 20:48:16.6 & +34:27:23 & B1 & 4.42 & $0.172\pm{0.018}$ & 1.235 & $4.72^{+0.35}_{-0.37}$ & $4410^{+670}_{-660}$ & 1 \\
2193902559325301760 & IRAS 20490+5934 & 20:50:13.4 & +59:45:50 & A3e & 3.95 & $2.062\pm{0.016}$ & 0.8655 & $0.4819^{+0.0035}_{-0.0037}$ & $151.4^{+2.2}_{-2.3}$ & 2 \\
2197298400984984064 & IRAS 20559+6416 & 20:56:53.5 & +64:28:32 &  &  & $2.241\pm{0.029}$ & 1.813 & $0.4394^{+0.0058}_{-0.0063}$ & $50.0^{+1.3}_{-1.4}$ & 2 \\
1731164844433296128 & IRAS 20547+0247 & 20:57:16.3 & +02:58:43 &  &  & $0.189\pm{0.054}$ & 3.259 & $4.8^{+1.1}_{-0.9}$ & $12300^{+6200}_{-4300}$ & 2 \\
2168803045330976768 & IRAS 20572+4919 & 20:58:55.6 & +49:31:15 & F3Ie & 3.83 & $1.577\pm{0.012}$ & 0.9983 & $0.6230^{+0.0052}_{-0.0038}$ & $118.4^{+2.0}_{-1.5}$ & 2 \\
2179471159976791168 & IRAS 21289+5815 & 21:30:22.8 & +58:28:52 & A9 & 3.88 & $1.011\pm{0.02}$ & 1.362 & $0.964^{+0.017}_{-0.017}$ & $22.33^{+0.78}_{-0.77}$ & 2 \\
6831062200578042624 & PHL 1580 & 21:30:25.2 & -19:22:33 &  &  & $3.156\pm{0.018}$ & 0.9539 & $0.3142^{+0.0016}_{-0.0015}$ & $2.369^{+0.024}_{-0.022}$ & 1 \\
6616993471402951680 & BPS CS29493-0046 & 21:50:05.8 & -30:49:11 & B & 4.26 & $0.05\pm{0.04}$ & 0.9237 & $14.8^{+4.6}_{-6.5}$ & $620^{+440}_{-420}$ & 2 \\
6824212243136821248 & PHL 174 & 21:50:48.6 & -19:42:00 & B3 & 4.26 & $0.063\pm{0.035}$ & 1.022 & $8.4^{+2.0}_{-1.7}$ & $600^{+320}_{-210}$ & 1 \\
1976077657917533952 & IRAS 21546+4721 & 21:56:33.1 & +47:36:13 &  &  & $0.028\pm{0.028}$ & 1.818 & $9.8^{+1.8}_{-1.5}$ & $2010^{+800}_{-580}$ & 1 \\
2005246464463628800 & IRAS 22023+5249 & 22:04:12.2 & +53:03:59 & B1 & 4.42 & $0.173\pm{0.018}$ & 1.22 & $5.28^{+0.58}_{-0.48}$ & $4300^{+1000}_{-700}$ & 1 \\
1958757291756223104 & IRAS 22223+4327 & 22:24:31.3 & +43:43:12 & F7I & 3.79 & $0.332\pm{0.026}$ & 1.685 & $2.68^{+0.20}_{-0.13}$ & $2340^{+360}_{-220}$ & 1 \\
2006425553228658816 & IRAS 22272+5435 & 22:29:10.4 & +54:51:07 & G5Ia & 3.7 & $0.686\pm{0.028}$ & 1.184 & $1.410^{+0.055}_{-0.054}$ & $6170^{+490}_{-460}$ & 1 \\
2594762641717531008 & IRAS 22327-1731 & 22:35:27.5 & -17:15:26 & A0III & 3.99 & $1.385\pm{0.056}$ & 2.313 & $0.702^{+0.028}_{-0.022}$ & $182^{+15}_{-11}$ & 1 \\
6385794694664872320 & CD-68 2300 & 22:40:48.1 & -67:41:18 & B8 & 4.06 & $0.704\pm{0.022}$ & 0.9883 & $1.390^{+0.049}_{-0.046}$ & $2470^{+180}_{-160}$ & 2 \\
1932229409071269248 & BD+39 4926 & 22:46:11.2 & +40:06:26 & B8 & 4.06 & $0.18\pm{0.037}$ & 2.188 & $4.8^{+1.0}_{-0.7}$ & $4000^{+1800}_{-1200}$ & 1 \\
2015785313459952128 & IRAS 23304+6147 & 23:32:44.7 & +62:03:51 & G2Ia & 3.73 & $0.237\pm{0.028}$ & 1.586 & $3.98^{+0.41}_{-0.36}$ & $5200^{+1100}_{-900}$ & 1 \\
\hline
\end{longtable}
\end{landscape}

\end{document}